\documentclass[twocolumn]{aastex62}

\usepackage{apjfonts}
\usepackage{graphicx}
\usepackage{wrapfig}
\usepackage{amsmath}
\usepackage{multirow}
\usepackage{booktabs}
\usepackage{float}

\newcommand{\Msun}{$M_{\sun}$}

\newcommand{\solperpc}{$M_{\sun}$ pc$^{-2}$}

\newcommand{\SigGas}{$\Sigma_{\mathrm{gas}}$}
\newcommand{\SigSFR}{$\Sigma_{\mathrm{SFR}}$}
\newcommand{\CSigSFR}{$\langle \Sigma_{\mathrm{SFR}} \rangle$}

\shorttitle{PHATTER. III. M33 Cluster Mass Function}
\shortauthors{Wainer et al.}

\begin{document}

\title{The Panchromatic Hubble Andromeda Treasury: Triangulum Extended Region (PHATTER). \\
III. The Mass Function of Young Star Clusters in M33}

\author[0000-0001-6320-2230]{Tobin M. Wainer}
\affiliation{Department of Physics and Astronomy, University of Utah, Salt Lake City, UT 84112, USA}
\affiliation{Center for Interdisciplinary Exploration and Research in Astrophysics (CIERA) and Department of Physics and Astronomy, Northwestern University, 1800 Sherman Ave., Evanston, IL 60201, USA}

\author[0000-0001-6421-0953]{L. Clifton Johnson}
\affiliation{Center for Interdisciplinary Exploration and Research in Astrophysics (CIERA) and Department of Physics and Astronomy, Northwestern University, 1800 Sherman Ave., Evanston, IL 60201, USA}

\author[0000-0003-0248-5470]{Anil C. Seth}
\affiliation{Department of Physics and Astronomy, University of Utah, Salt Lake City, UT 84112, USA}

\author[0000-0001-9961-8203]{Estephani E. TorresVillanueva}
\affiliation{Department of Physics and Astronomy, 
University of Utah, Salt Lake City, UT 84112, USA}
\affiliation{Department of Astronomy, University of Wisconsin-Madison, Madison, WI, 53706, USA}

\author[0000-0002-1264-2006]{Julianne J. Dalcanton}
\affiliation{Department of Astronomy, University of Washington, Box 351580, Seattle, WA 98195, USA}
\affiliation{Center for Computational Astrophysics, Flatiron Institute, 162 Fifth Avenue, New York, NY 10010, USA}

\author[0000-0001-7531-9815]{Meredith J. Durbin}
\affiliation{Department of Astronomy, University of Washington, Box 351580, Seattle, WA 98195, USA}

\author[0000-0001-8416-4093]{Andrew Dolphin}
\affiliation{Raytheon, Tucson, AZ 85726, USA}
\affiliation{Steward Observatory, University of Arizona, Tucson, AZ 85726, USA}

\author[0000-0002-6442-6030]{Daniel R. Weisz}
\affiliation{Department of Astronomy, University of California Berkeley, Berkeley, CA 94720, USA}

\author[0000-0002-7502-0597]{Benjamin F. Williams}
\affiliation{Department of Astronomy, University of Washington, Box 351580, Seattle, WA 98195, USA}

\author{PHATTER Collaboration}
\noaffiliation

\correspondingauthor{Tobin M. Wainer}
\email{tobin.wainer@utah.edu}

\begin{abstract}
    We measure the star cluster mass function for the Local Group galaxy M33. We use the catalog of stellar clusters selected from the Panchromatic Hubble Andromeda Treasury: Triangulum Extended Region (PHATTER) survey. We analyze 711 clusters in M33 with $\rm 7.0 <  log(Age/yr) < 8.5$, and log($M/M_{\odot}$) $>$ 3.0 as determined from color-magnitude diagram fits to individual stars. The M33 cluster mass function is best described by a Schechter function with power law slope $\alpha = -2.06^{+0.14}_{-0.13}$, and truncation mass log($M_c/M_{\odot}$) $= 4.24^{+0.16}_{-0.13}$. The data show strong evidence for a high-mass truncation, thus strongly favoring a Schechter function fit over a pure power law. M33's truncation mass is consistent with the previously identified linear trend between $M_c$, and star formation rate surface density, \SigSFR. We also explore the effect that individual cluster mass uncertainties have on derived mass function parameters, and find evidence to suggest that large cluster mass uncertainties have the potential to bias the truncation mass of fitted mass functions on the one sigma level.
\end{abstract}

\keywords{Star clusters (1567), Star formation (1569), Triangulum Galaxy (1712)}

\section{Introduction}

Star clusters are fundamental probes of galaxy evolution and the process of star formation. Ancient globular clusters trace the halos and formation histories of galaxies \citep[e.g.][]{kruijssen_formation_2019}, while young clusters trace the quantity and characteristics of on-going and recent star formation.  For young clusters, studies have shown a correlation between the star formation rate (SFR) surface density, \SigSFR, and the fraction of stars that form in clusters \citep{larsen_mass_2009,johnson_panchromatic_2016, adamo_legacy_2017}. More intense star formation leads to a higher fraction of a galaxy’s stars being formed in star clusters. Because of this link, star clusters encode a record of galactic star formation activity in their population characteristics \citep[e.g.][]{johnson_panchromatic_2017}. Furthermore, studying the properties of star cluster formation as a function of galactic environment can help us better understand the star formation process -- for example, regarding the efficiency of star formation and the role of stellar feedback \citep[e.g.,][]{grudic_model_2021}.

One measurable trait of a star cluster population is the cluster mass function (CMF). The mass function for young star clusters has been observed to be consistent with a power law distribution ($dN / dM \propto M^{\alpha}$) where $\alpha \sim -2$  \citep{portegies_zwart_young_2010, krumholz_star_2019}; this is similar to the giant molecular cloud mass function, but due to hierarchical cluster growth, mapping between the two mass functions is difficult \citep[e.g ][]{longmore_formation_2014}. The measurement of CMFs is complicated by a number of factors. First, different mass ranges of clusters are studied in different galaxies \citep[e.g.][]{bik_clusters_2003,zhang_mass_1999, larsen_mass_2009}. Second, different methods for identifying, and measuring cluster ages and masses can significantly impact CMF measurements \citep[see recent review][]{krumholz_star_2019}.  

Initially, the observed power law slopes of CMFs suggested that there may be a universal power law CMF. However, increasing evidence suggests that the masses of young clusters deviates from a power law at high masses, and instead young star clusters follow a \cite{schechter_analytic_1976} function with an exponential truncation \citep[$dN / dM \propto M^{\alpha} \exp(-M / M_c)$;][]{larsen_mass_2009,bastian_stellar_2012,adamo_probing_2015,johnson_panchromatic_2017,adamo_legacy_2017, lieberz_origin_2017, messa_young_2018}.

Constraining the Schechter truncation mass can be difficult due to small number statistics of massive clusters and small predicted differences between Schechter and non-truncated power law models. Some studies continue to favor a power law model or very large Schechter truncation masses \citep{chandar_age_2016, cook_star_2019, mok_constraints_2019, whitmore_legus_2020}. However, a growing number of truncation detections using high-quality cluster data and strong statistical fitting techniques anchor a growing body of evidence favoring a Schechter CMF. The $M_c$ of $8.5 \times 10^3$ \Msun\ determination in M31 made using a sample of 840 clusters with ages between 10-300~Myr \citep{johnson_panchromatic_2017} provides particularly convincing evidence in favor of a truncation \citep{krumholz_star_2019}. 

Like the fraction of stars that form in clusters, the truncation of the CMF also appears to vary with the intensity of star formation. In the observations cited above, clusters in high intensity star formation environments follow power law mass distribution extending up to $\sim$10$^6$~M$_\odot$, while clusters in more normal star-forming galaxies have CMFs with high-mass truncations at $\sim$10$^5$~M$_\odot$. The truncation masses are even lower in relatively quiescent galaxies, which form very few high mass clusters. Using a handful of measurements with reliable truncation detections, \citet{johnson_panchromatic_2017} find that there is a nearly linear relationship between \SigSFR\ and $M_c$, where $M_c \propto \langle \Sigma_{\mathrm{SFR}} \rangle ^{1.1}$.

A trend between the maximum star cluster mass scale and a galaxy's star formation properties is not surprising. High \SigSFR\ is physically connected to increasing \SigGas\ and pressure, both of which are associated with increased star formation efficiency, cluster formation efficiency (or bound stellar fraction), and resulting stellar density \citep[see e.g.,][]{elmegreen_pressure_2009, kruijssen_fraction_2012, reina-campos_unified_2017, elmegreen_two_2018, grudic_model_2021}. While the development of theoretical models to predict and understand the maximum cluster mass scale are still on-going, there is increasing consensus that variations and trends like the ones we observe are expected given our current understanding of star formation. However, we are motivated to confirm and quantify such trends to further constrain the process of cluster formation.
    
One notable strength of the M31 CMF study was its use of high-precision ages and masses that were inferred from resolved color-magnitude diagrams (CMDs) of member stars in each cluster. This methodology is not possible in more distant galaxies, making studies of Local Group galaxies particularly valuable. Beyond the Local Group however, cluster ages and masses can only be inferred from integrated light fitting methods. These methods for determining cluster properties have been proven to be less reliable than traditional CMD isochrone fitting \citep{krumholz_star_2019, johnson_catalog_2022}. It is imperative to capitalize on the small number of Local Group galaxies, such as M33, where high-precision samples are accessible and obtain high quality cluster mass function determinations for these targets.

In this work, we measure and analyze the CMF for the Local Group galaxy M33. Our work utilizes data from the Panchromatic Hubble Andromeda Treasury: Triangulum Extended Region (PHATTER) survey detailed in \citet{williams_panchromatic_2021}, and the cluster sample from \citet{johnson_catalog_2022}. We determine cluster ages and masses through maximum-likelihood CMD analysis, and we fit the CMF using a probabilistic Bayesian approach.
Although M33's \SigSFR\ is somewhat higher than that of M31, star formation in M33 is relatively quiescent compared to other galaxies with previous CMF measurements. Therefore, M33 occupies a valuable place in parameter space in the investigation of CMF truncation behavior.
    
We structure the paper as follows. First we present the data in Section~\ref{sec: data}, and then lay out the probabilistic approach for fitting the cluster mass function in Section~\ref{sec: analysis}. We will then present the CMF fitting results Section~\ref{sec: results}. In Section~\ref{sec: discusion}, we will compare our CMF results to other galaxies' published CMFs and further analyze the link between a galaxy's CMF and \SigSFR.  We also examine the implications of mass measurement uncertainties on CMF fitting results, and show that high cluster mass uncertainties can bias $M_c$ measurements towards high values.

\section{Data}\label{sec: data}

Our cluster sample is drawn from the Local Group Cluster Search (LGCS) cluster catalog \citep{johnson_catalog_2022}, which was created using data from the PHATTER survey \citep{williams_panchromatic_2021}. PHATTER uses the \textit{Hubble Space Telescope} (HST) Advanced Camera for Surveys (ACS) and Wide Field Camera 3 (WFC3) to image the inner region of M33's disk in six filters from UV to IR. The \citet{johnson_catalog_2022} catalog presents 1214 star clusters identified through a crowdsourced, visual search of the optical PHATTER data (F475W \& F814W), facilitated by the LGCS project\footnote{\url{https://www.clustersearch.org}}, a citizen science effort hosted on the Zooniverse\footnote{\url{https://www.zooniverse.org}} platform. 
We adopt the recommended cluster catalog threshold ($f_{\textrm{cluster, W}} > 0.674$), which limits the contamination rate to $\lesssim$4\% for our cluster sample.

\subsection{Derivation of Cluster Ages and Masses}

For each cluster in our sample, we extract an optical color magnitude diagram (CMD) in the F475W and F814W filters. These passbands yield the deepest CMDs available from the PHATTER data. Specifically, we use CMDs composed of stars which lie within the photometric aperture ($R_{ap}$) derived in \citet{johnson_catalog_2022}, which corresponds to approximately three times the cluster half-light radius. We assume all stars within $R_{ap}$ are cluster members, and all members are within $R_{ap}$. We make no correction for mass that lies outside $R_{ap}$. Based on experiments with synthetic clusters, we expect $<$30\% of light to fall outside our photometric aperture. The median aperture correction for clusters in our sample is -0.04 mags, with the largest correction being -0.69. This suggests we may be losing $\sim$4\% of the light from our clusters in typical cases, and thus underestimating clusters logM by 0.02 dex.  This number is much smaller than the uncertainties in our $M_c$ values derived below, and we don’t correct our masses for this aperture correction. We characterize the surrounding field star population using an annulus which spans 1.2--3.4 $R_{ap}$. This annulus has 10$\times$ the area of the cluster aperture. The background is fit along with the cluster models; the scaling of the background is a free parameter in the fit.

We use the MATCH software package to perform maximum-likelihood CMD fits and derive constraints on cluster properties following techniques described in \citet{dolphin_numerical_2002}. Unlike its typical use in determining time-resolved star formation histories, we use MATCH in a more limited simple stellar population (SSP) mode. Here, the model CMD is composed of a population drawn from a single time bin, rather than a linear combination of populations from multiple time bins.

The MATCH code uses theoretical isochrones to populate synthetic CMDs according to input parameters of age, dust extinction ($A_V$), distance, metallicity, stellar initial mass function (IMF), and binary fraction. Synthetic populations are created from unique combinations of input parameters, convolved with a model of observational completeness and noise derived from artificial star tests (ASTs), to produce a simulated CMD. This simulated CMD is combined with a background model created from the CMD of stars lying in an annulus surrounding the cluster, then both the model and background components are scaled to best reproduce the observed cluster CMD. The fit quality is calculated according to a Poisson likelihood statistic and the code iterates through a grid of input parameter value combinations to map the distribution of probability. 

We adopt the M33 distance modulus of 24.67 \citep{de_grijs_clustering_2014}, and assume a metallicity $[M/H]$ of  $-0.15 \pm 0.25$ based on the range of present-day gas phase metallicity in M33 \citep[e.g.][]{u_new_2009}. For young clusters, the age is heavily weighted by the main sequence and thus has little metallicity sensitivity. For consistency with previous studies, our assumptions in stellar modeling follow the cluster fitting of \citet{weisz_high-mass_2015, johnson_panchromatic_2016}. Briefly, we adopt a binary fraction of 0.35 with a uniform mass ratio distribution, a \citet{kroupa_variation_2001} stellar IMF between 0.15--120~\Msun, and Padova stellar models \citep{marigo_evolution_2008} with low mass asymptotic giant branch tracks from \citet{girardi_acs_2010}.

For each cluster we perform 25,000 ASTs to ensure accurate characterization of photometric completeness and noise, encompassing a wide range of CMD positions and cluster radii. Input positions for ASTs are distributed based on the measured half-light radii of the clusters, and assume a \citet{king_structure_1962} profile with a concentration of 10.

We compute CMD fits for a grid of age (6.6 $<$ log(Age/yr) $<$ 9.0) and dust extinction (0 mag $< A_V < 2$ mag), deriving the mass at each grid point from the best fit CMD model. We use relative likelihoods derived across the grid to obtain marginalized probability distribution functions (PDFs) for each parameter. We adopt the best fit model for mass, age, and $A_V$, and assign each an uncertainty defined by the 16th and 84th percentile of the PDF.  The masses derived represent the initial masses for the clusters. 

\begin{figure}
    \centering
    \includegraphics[width=8.5cm]{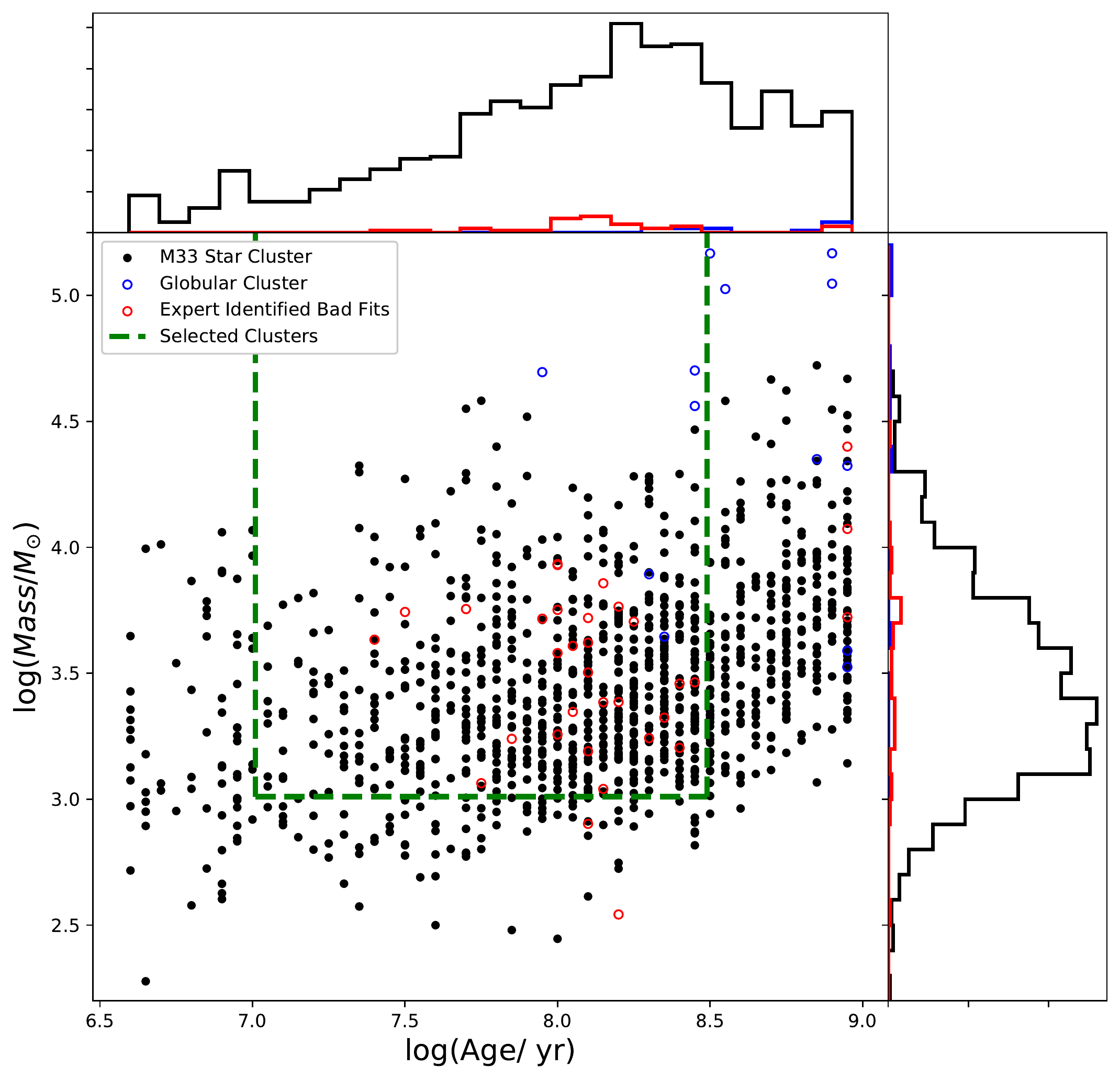}
    \caption{M33 cluster age and mass estimates for the \citet{johnson_catalog_2022} catalog. Black points represent clusters with good age and mass estimates.  Objects unanimously identified by a group of co-authors to be globular clusters with true ages that are much older than the CMD fits are shown in blue, while other bad CMD fits are shown in red (see text for details). The green dashed lines denote the sample selection for the 711 clusters we use for CMF fitting as discussed in Section \ref{ssec:cluster_selection}. Additional panels show the one-dimensional distributions for mass and age.}
    \label{fig:age_vs_mass}
\end{figure}

Figure~\ref{fig:age_vs_mass} shows the full sample of best-fit ages and masses for the M33 cluster sample. We find the majority of the cluster age and mass PDFs to be 
Gaussian.  In $<10\%$ of cases the PDFs are bi-modal or have a tail to lower or higher values. The median 1$\sigma$ uncertainty in age is 0.11 dex, while for mass the median is 0.04 dex. The full catalog of cluster parameters is presented in Table \ref{tab:cmd_est}. For the remainder of the paper, we use the maximum a posteriori values for the cluster ages and masses; these are contained in the Best-fit column in Table \ref{tab:cmd_est}.

While our results focus only on clusters $<$300~Myr, we note that there also exists of a relatively large number of massive $\sim$1~Gyr old clusters. A similar sample of clusters is not present in the M31 PHAT data \citep{johnson_panchromatic_2016}.

\begin{deluxetable*}{ccc|cccc|cccc|cccc}
\tabletypesize{\footnotesize}
\setlength{\tabcolsep}{0.05in}
\tablecaption{CMD Property Estimates} \label{tab:cmd_est}
\tablewidth{0pt}
\tablehead{
\colhead{ID} & \colhead{$N_{MS}$} & \colhead{Exclude Flag} & \multicolumn{4}{|c|}{log($Age/yr$)} & \multicolumn{4}{c|}{log($Mass/M_{\odot}$)} & \multicolumn{4}{c}{$A_{v}$}  \\
\colhead{} & \colhead{} & \colhead{} & 
\multicolumn{1}{|c}{P16} & \colhead{P50} & \colhead{P84} & \colhead{Best-fit} & \multicolumn{1}{|c}{P16} & \colhead{P50} & \colhead{P84} & \colhead{Best-fit} & \multicolumn{1}{|c}{P16} & \colhead{P50} & \colhead{P84} & \colhead{Best-fit}
}

\startdata
1 & 293 & 0.0 & 8.01 & 8.04 & 8.08 & 8.0 & 3.76 & 3.77 & 3.79 & 3.76 & 0.11 & 0.12 & 0.14 & 0.1 \\
2 & 796 & 0.0 & 8.12 & 8.22 & 8.29 & 8.1 & 3.56 & 3.59 & 3.63 & 3.56 & 0.30 & 0.34 & 0.43 & 0.4 \\
3 & 796 & 0.0 & 8.34 & 8.38 & 8.42 & 8.4 & 3.41 & 3.43 & 3.46 & 3.45 & 0.26 & 0.29 & 0.34 & 0.3 \\
4 & 530 & 0.0 & 8.91 & 8.92 & 8.94 & 8.95 & 4.05 & 4.09 & 4.13 & 4.09 & 0.12 & 0.24 & 0.30 & 0.2 \\
5 & 587 & 0.0 & 8.31 & 8.33 & 8.34 & 8.3 & 4.16 & 4.18 & 4.21 & 4.19 & 0.56 & 0.58 & 0.66 & 0.6 \\
\enddata

\tablecomments{Table \ref{tab:cmd_est} is published in its entirety in machine-readable format. A portion is shown here for guidance regarding its form and content.  $N_{MS}$ gives the number of bright main sequence stars in the subimage with the cluster, while the Exclude Flag is set to one for clusters with bad CMD properties (see text for more details).  The rest of the columns provide the results of our CMD fitting, including 16th, 50th and 84th percentiles of the marginalized 1D PDFs for each parameter, as well as the best fit parameter values. }

\end{deluxetable*}

\subsection{Cluster Selection for Mass Function Analysis} \label{ssec:cluster_selection}

After fitting the cluster CMD's, we perform a visual inspection of each cluster's CMD fits and optical images. From this inspection, we exclude results for 33 clusters that a group of co-authors unanimously agreed were poor fits. These clusters are older clusters with few detected member stars that were poorly and erroneously fit with a young, high $A_V$ model. These clusters are denoted by an `exclude' flag in Table \ref{tab:cmd_est}, and are represented by the red open circles in Figure \ref{fig:age_vs_mass}. We note that excluding these flagged fits does not significantly impact the CMF results (i.e., differences in parameter fits with and without these clusters are much less than $1\sigma$). We also visually identify 13 globular cluster candidates (blue open circles in Figure \ref{fig:age_vs_mass}) and exclude these objects from our sample; globular clusters are known failure cases for our CMD analysis due to limits in the age and metallicity range assumed for the fitting. Most (11/13) of these candidates appear in previous catalogs \citep{san_roman_photometric_2010}, and every cluster in this subsample with an existing age estimate from \citet{fan_star_2014} was reported as $>1$ Gyr old.  

We select a sample of clusters for mass function analysis in the age range $7.0 < {\rm log(Age/yr)} < 8.5$. The lower threshold is adopted because during the first 10 Myr of a clusters life, clusters are embedded, making optical observations difficult, and our sample incomplete. In addition, young embedded groupings of stars (< 10 Myr), are still forming through hierarchical merging of sub-clumps, making it unclear whether the resulting structure will be a long-lived, gravitationally bound cluster \citep{allison_early_2010, gieles_all_2012, messa_looking_2021}. The upper threshold is where CMD fitting becomes less accurate due to the cluster's main sequence turnoff dropping below the 50\% detection limit for the stellar photometry.

\begin{figure*}
    \centering
    \includegraphics[width=\textwidth]{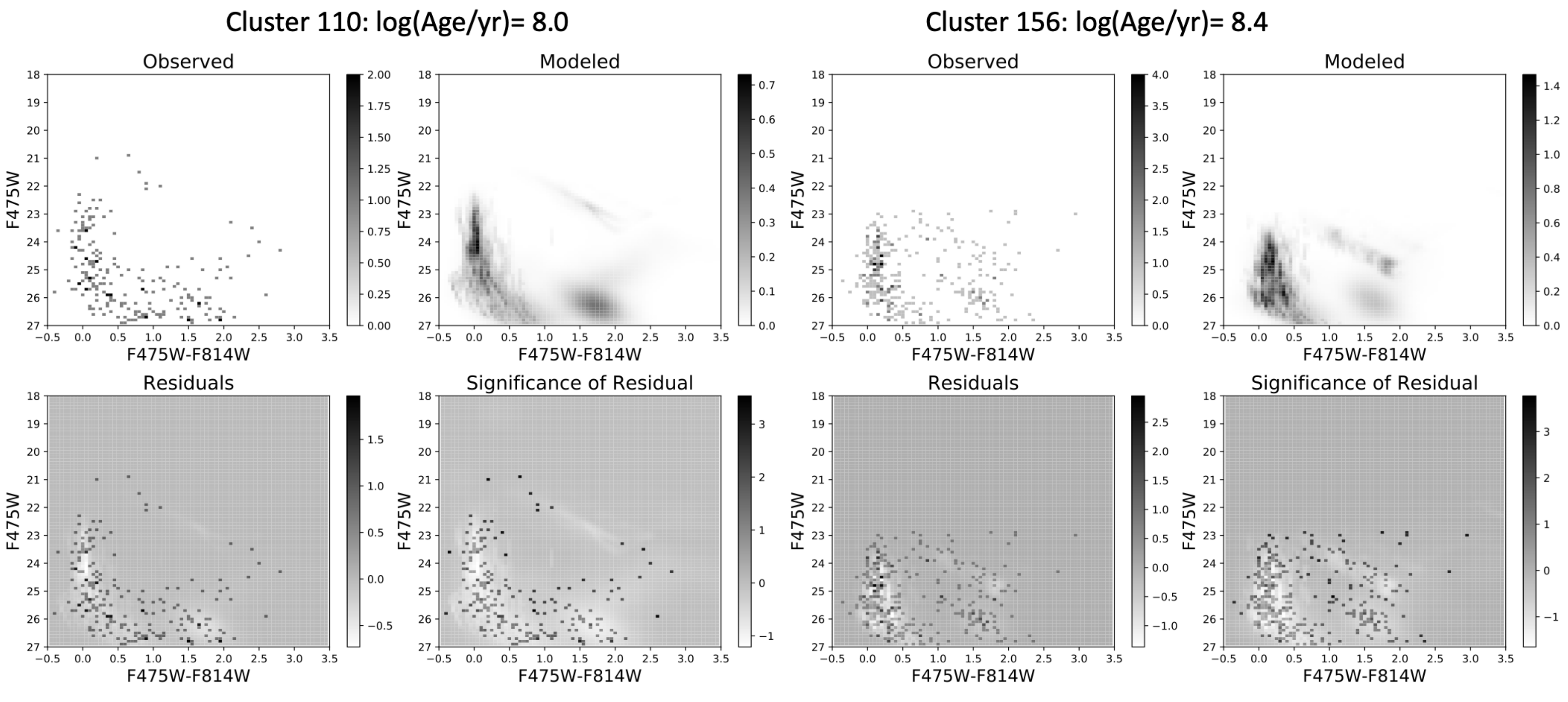}
    \caption{Quality assessment of CMD fits to two clusters.  Panels show the observed F475W-F814W vs F475W CMD, modeled CMD, residuals, and significance of the residuals for cluster IDs 110 (left four panels), and 156 (right four panels). Cluster ID 110 has a best fit age estimate of 100 Myrs, representing the median age of our final sample, while cluster ID 156 shows a somewhat older cluster showing that good age estimates are possible out to 300~Myr.}
    \label{fig:residuals}
\end{figure*}

In Figure~\ref{fig:residuals}, we present CMD fitting results for two example clusters that lie at the median age ($\sim$100 Myr; LGCS-M33 110) and maximum age ($\sim$300 Myr; LGCS-M33 156) of the cluster sample.  For each cluster, we show the observed CMD, modeled CMD, residuals, and the significance of the residuals. 

We find that the uncertainty of our age estimates increase as cluster masses decrease, and that large age uncertainties also translate to less reliable mass estimates. As a result, we adopt a minimum mass of 1000 \Msun\ for our CMF fitting cluster sample. Below this mass limit, the average $1\sigma$ error in log(Age/yr) is $0.30$ dex for cluster masses from $2.8 \leq \log(M/M_{\odot}) < 3.0$, where as the average $1\sigma$ error improves to 0.21 dex for clusters with masses from $3.0 \leq \log(M/M_{\odot}) < 3.2$. We note that this selection of minimum mass has a less than 1$\sigma$ variance in the inferred CMF.

\begin{figure}
    \centering
    \includegraphics[width=0.47\textwidth]{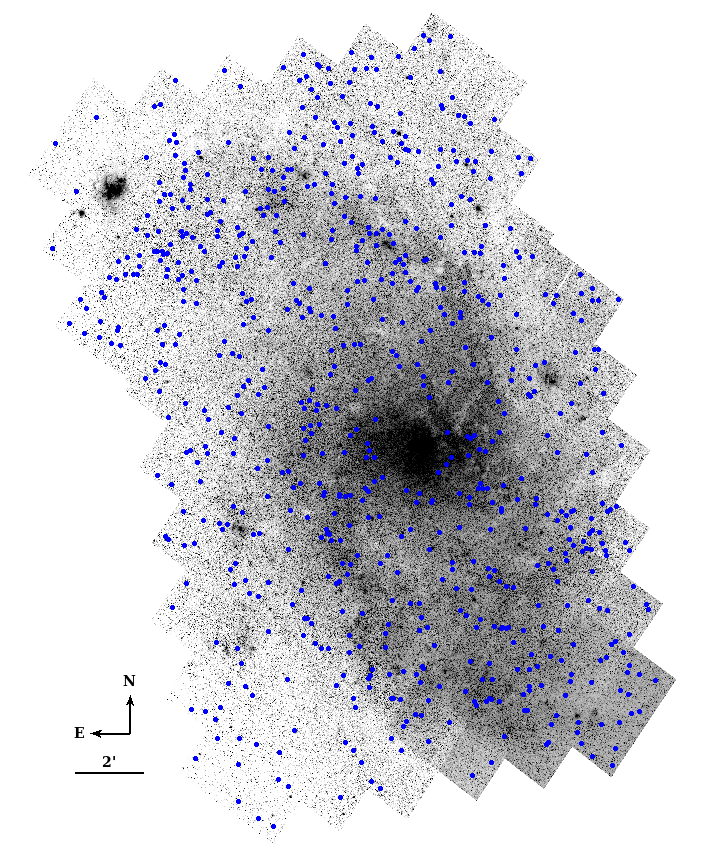}
    \caption{Spatial distribution of the analyzed 711 clusters with $\rm 7.0 <  log(Age/yr) < 8.5$, and log($M/M_{\odot}$) $>$ 3.0 overlaid on the PHATTER F475W image.}
    \label{fig:map}
\end{figure}

Our final cluster sample for CMF fitting includes 711 clusters. The selected age and mass range is denoted by the green dashed lines in Figure \ref{fig:age_vs_mass}, and the spatial distribution of the selected sample is shown in Figure \ref{fig:map}. For this selected sample of clusters, the median 1$\sigma$ age uncertainty is 0.10 dex and the median 1$\sigma$ mass uncertainty is 0.03 dex. The distribution of cluster mass errors is discussed further in Section~\ref{sec: discusion}.

\subsection{Cluster Sample Completeness} \label{ssec: completeness}

Once we obtained measurements of cluster ages and masses, we needed to correct the cluster catalog for completeness to properly fit the intrinsic mass distribution. The full description of the cluster sample completeness can be found in Section 4 of \citep{johnson_catalog_2022}. Briefly, the completeness of the cluster sample was determined by measuring the detections of synthetic clusters placed in LGCS images \citep[Section 2.4 of][]{johnson_catalog_2022}. The synthetic clusters are characterized by log(Age/yr) vs. log(Mass/M$_\odot$), binned as a function of log(Age/yr). The completeness as function of log(M/M$_\odot$), $C$, is characterized using the functional form of a logistic function given by:
\begin{equation} 
    C(M_{50}, {\rm log(M/M_\odot)}) = ( 1 + exp[ -k ( {\rm log(M/M_\odot)} - M_{50}) ] )^{-1}
\end{equation} 
where $k$ sets the slope of the logistic function (fixed to 6.02 as explained in \citet{johnson_catalog_2022}) and $M_{50}$ is the 50\% completeness limit. $M_{50}$ is well described by an exponential function:
\begin{equation} \label{eq: comp_from_cat_pap}
    M_{50}(\tau) = a \times exp((b\times (\tau -\tau_{min}))) + c
\end{equation}
where $\tau \equiv {\rm log(Age/yr)}$ and $\tau_{min}$ is the median cluster log(Age/yr) in the youngest bin, which is 7.09. The constants $a$, $b$, and $c$ were fit through minimizing the $\chi^2$ on the binned data set of synthetic cluster detections over the age range of $7.0 < {\rm log(Age/yr)} < 8.5$. The best fit $M_{50}(\tau)$ parameters are $a=0.0303, b=1.9899, c=2.9770$, with a reduced $\chi^2$ of 0.82.

\citet{johnson_catalog_2022} finds environment plays a significant role in completeness, largely due to the density of main sequence stars in search fields that contain a cluster. We characterize the number of main sequence stars per LGCS search image ($\sim$36$\times$25 arcsec) as log($N_{MS}$), which ranges from $2.0 < \log(N_{MS}) < 3.75$ in M33. \citet{johnson_catalog_2022} finds that over this range, the 50\% mass completeness is impacted by up to 0.6 dex. 

\begin{figure}
    \centering
    \includegraphics[width=8.5cm]{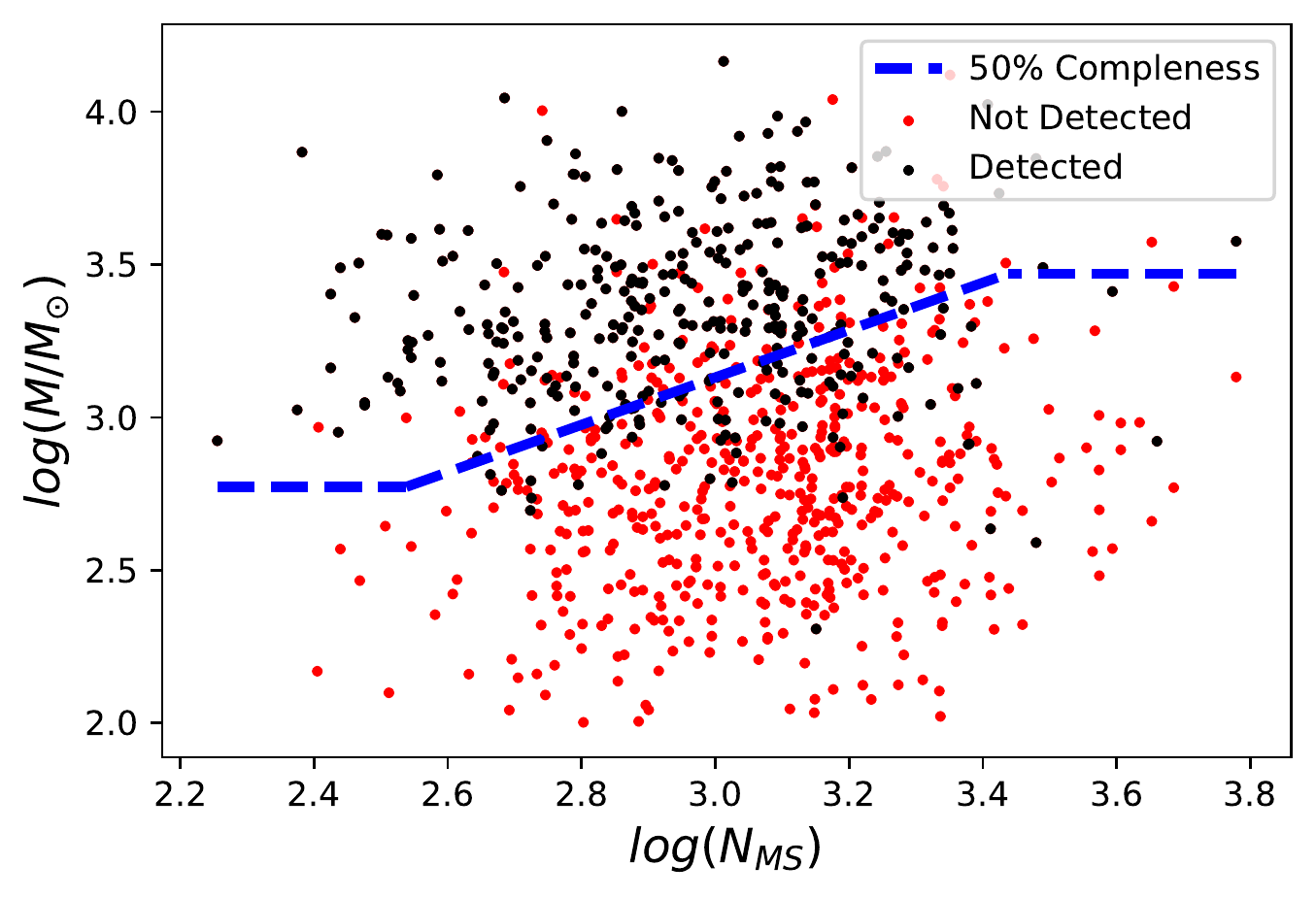}
    \caption{Completeness results as a function of environment from our synthetic cluster analysis. The effect of environment is quantified based on the number density of bright main sequence stars quantified as log($N_{MS}$). Black points show synthetic clusters that were detected by citizen scientists in \citet{johnson_catalog_2022}, while red were not detected. The dashed blue line shows the 50\% mass completeness as a function of log($N_{MS})$, which follows Equation \ref{eq:nms_dependence}.}
    \label{fig:NMS_com}
\end{figure}

We further examine the environmental completeness dependence here and incorporate it into our completeness model, by deriving a relationship between the 50\% mass completeness and log($N_{MS}$). We first exclude synthetic clusters that are greater than $2\sigma$ outliers in log($N_{MS}$), then split the cluster into three equal bins of $\log(N_{MS})$, and fit the mass completeness as a function of log(Age/yr) for each bin. We find a linear trend between the average 50\% mass completeness values for median values of the three $\log(N_{MS})$ bins, which we characterize using a linear function $m \times (\log(N_{MS})) + b_{nms}$, where $m=0.7727$ and $ b_{nms}=0.6674$. We adopt the value at the bin edge for $\log(N_{MS})$ values lower than the lowest bin and higher than the edge of the highest bin to avoid extrapolating where our number statistics are minimal. The 50\% mass completeness as a function of log($N_{MS}$) is shown in Figure \ref{fig:NMS_com}. 

We incorporate the environmental impact on completeness from Figure~\ref{fig:NMS_com} into the sample completeness function by having the $c$ parameter in Equation~\ref{eq: comp_from_cat_pap} be dependent on log($N_{MS}$), such that our completeness function becomes,
\begin{equation} \label{eq:50_completeness}
    M_{50}(\tau, N_{MS}) = a \times exp (b(\tau -\tau_{min})) + c(N_{MS})
\end{equation}
and $c(N_{MS})$ is described by,

\begin{equation} \label{eq:nms_dependence}
c(N_{MS}) = 
\begin{cases}
2.77
\hspace{112pt} log(N_{MS}) < 2.53 \\
( m \times log(N_{MS})+ b_{nms} ) 
\hspace{14pt} 2.53 \le log(N_{MS}) \le 3.49 \\
3.47
\hspace{112pt} log(N_{MS}) > 3.49
  
\end{cases}
\end{equation}

\begin{figure}
    \centering
    \includegraphics[width=8.5cm]{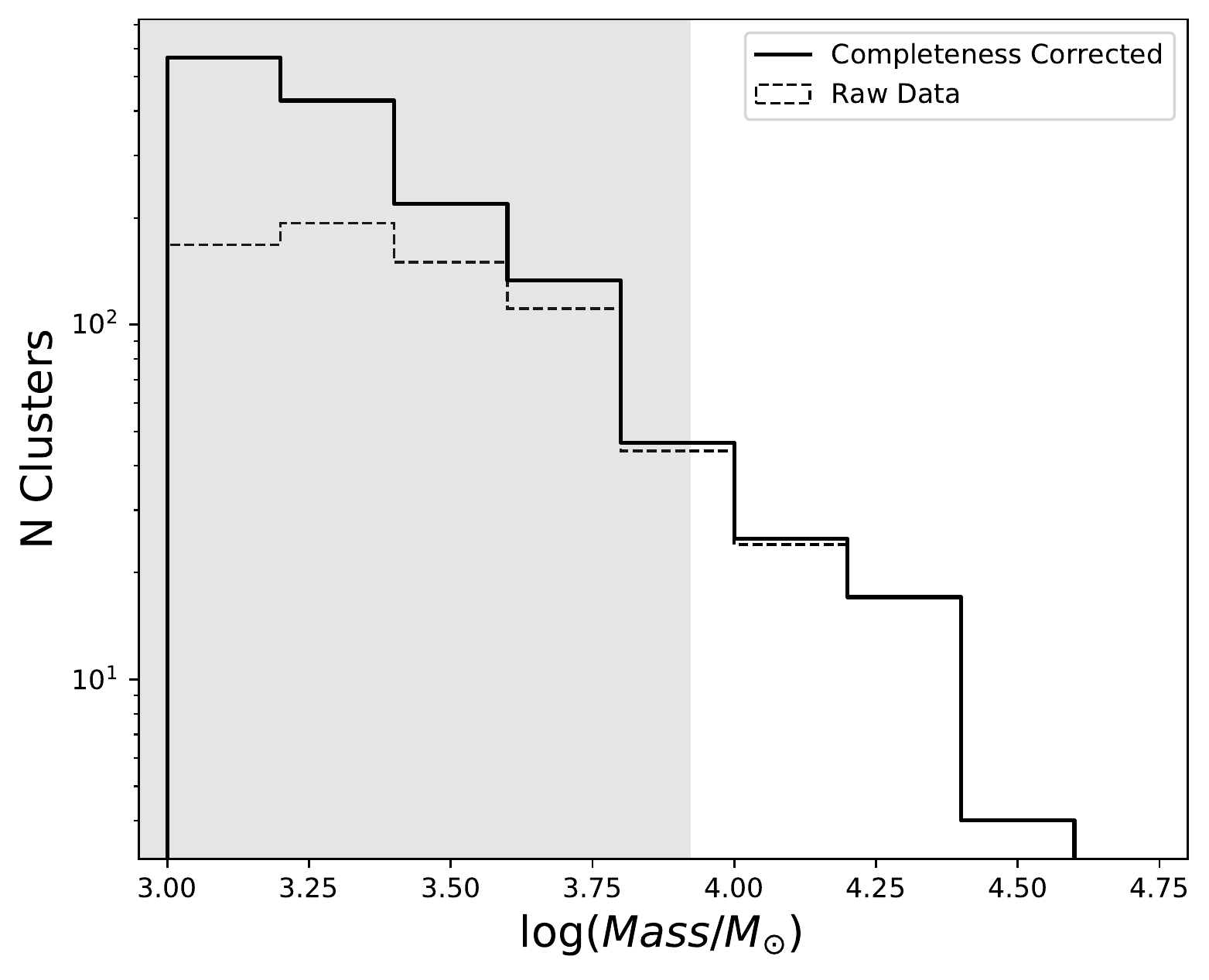}
    \caption{Observed and completeness corrected mass distributions for our cluster sample.  These mass distributions include clusters with mass above 10$^3$~M$_\odot$ and age between 10 and 300 Myr. We present raw number counts in the dashed lines and the completeness-corrected distribution in the solid lines. The gray region represents the range of 50\% completeness limits across our M33 sample.}
    \label{fig:binned_comp_cor}
\end{figure}

In Figure~\ref{fig:binned_comp_cor}, we show the completeness correction with respect to the raw data. We note that this figure is a visual representation of binned data, and the completeness correction is done on a cluster by cluster basis.  

\section{Probabilistic Analysis} \label{sec: analysis}

We follow the statistical methodology of \citet{johnson_panchromatic_2017} and perform probabilistic cluster mass function fitting. We deviate from \citet{johnson_panchromatic_2017} methodology only in the application of completeness as \citet{johnson_panchromatic_2017} evaluates the 50\% mass completeness for two age bins, while we improve this treatment to include the 50\% mass completeness for each individual cluster. Using the masses of all the clusters, we run the Markov Chain Monte Carlo (MCMC) code, \texttt{emcee} \citep{foreman-mackey_emcee_2013}, which takes advantage of the affine invariant ensemble sampler of \citet{goodman_ensemble_2010} to determine the functional form of the mass function through maximizing the likelihood of each cluster belonging to a mass function with given parameters. We then derive the posterior probability distributions of Schechter and power law mass function parameters.

For our MCMC calculation, we use 500 walkers, each performing 500 steps, of which we discard the first 100 burn-in steps. We ensure convergence of our chains according to the autocorrelation time, which we estimate to be 30 steps, far surpassed by our burn in period. For the power law function we report the median value of the marginalized posterior probability distribution function (PDF) for the power law slope parameter $\alpha$, as well as the $1\sigma$ confidence interval representing the 16th and 84th percentile range of the marginalized PDF. For the Schechter functional form we present the median value of the marginalized PDF for the two parameters, the power law slope ($\alpha$), and cluster mass cutoff ($M_c$), as well as the $1\sigma$ confidence interval for each parameter.

\subsection{Bayesian Approach}
The likelihood function for an observed cluster with mass $M$ is given as 
\begin{equation}
    p_{cluster}(M|\theta, \tau) \equiv \frac{1}{Z} p_{MF}(M| \theta)  p_{obs}(M| \tau),
\end{equation}
where the cluster distribution is represented by $p_{MF}(M| \theta)$ defined by parameters $\theta$, and $p_{obs}(M| \tau)$ is the observational completeness given as a function of cluster age $\tau$. In order to have the likelihood integrate to 1, the normalization $Z$ is required and given by
\begin{equation}
    Z=\int p_{MF}(M| \theta)  p_{obs}(M| \tau, N_{MS}) dM.
\end{equation}
We note that this normalization is calculated for each cluster, and does not refer to Bayesian evidence.

We test a pure power law distribution where the only parameter is given by $\theta = (\alpha$); $\alpha$ is defined as the power law index. We also adopt a \cite{schechter_analytic_1976} functional form that has two parameters given by $\theta= (\alpha, M_c)$ where $\alpha$ is the power law index for low-mass clusters, and $M_c$ is the characteristic mass defining the exponential high mass truncation. The \cite{schechter_analytic_1976} distribution follows the form:
\begin{equation}
    p_{MF}(M|\alpha, M_c) \propto (M/M_c)^{\alpha} exp(-M / M_c),
\end{equation}
as first used in the cluster context by \cite{larsen_mass_2009}.

To limit the impact of clusters with low completeness, we cut out any clusters below their local 50\% completeness limit, $M_{50}(\tau, N_{MS}$), as expressed in Equation \ref{eq:50_completeness}. Mathematically, this cluster selection can be described as:
\begin{equation}
p_{obs}(M | \tau, N_{MS}) =
\begin{cases}
\left(1+\exp\left[\frac{-k(M-M_{{50}}(\tau, N_{MS}))}{M_{\odot}} \right] \right)^{-1}, \\
\vspace{5mm} 
\hspace{90pt} M > M_{{50}}(\tau, N_{MS})\\
0, \\
\hspace{110pt} {otherwise}.
\end{cases}
\end{equation}

We find 533 clusters have masses larger than the local 50\% completeness limit.

We use Bayes' theorem to derive the posterior probability distribution function for the mass function parameters given as
\begin{equation}
    p(\theta|{M_i}, \tau) \propto p_{cluster}(\theta| {M_i}, \tau) p(\theta),
\end{equation}
where $p(\theta)$ is the prior probability of the Schechter parameters, ${M_i}$ is a set of 533 clusters, and $p_{cluster}(\theta| {M_i}, \tau)$ is the combined likelihood function for the full set of clusters defined below.  We adopt a uniform tophat prior for the Schechter parameters that covers the range of published values with plenty of cushion: $-3 \leq \alpha \leq -1$ and $3 \leq \log(M_c/M_{\sun}) \leq 8$.

The assumption with this implementation of a probabilistic approach is that the cluster mass uncertainties are negligible. \citet{weisz_panchromatic_2013} demonstrates that this assumption is valid if the fractional mass uncertainties are smaller than 10\%. However, when the fractional error of the masses extends to 50\% and beyond, fitting results can become affected \citep{johnson_panchromatic_2017}. We will further discuss this assumption in Section \ref{sec: discusion}.


The likelihood function for a set of clusters is defined as the product of the individual cluster mass probabilities given by, 
\begin{equation}
    p({M_i}| \theta, \tau) \equiv  \prod_{i=1}^{N} \frac{1}{Z_i} ({M_i}/M_c)^{\alpha} exp(-M_i / M_c) p_{obs}(M_i| \tau),
\end{equation}
where the normalization becomes
\begin{equation}
    Z_i = \int_{M_{50}}^{\infty}  ({M_i}/M_c)^{\alpha} exp(-M_i / M_c) p_{obs}(M_i| \tau, N_{MS}).
\end{equation}

\noindent{\em Power Law Model:} For comparison, we also fit for a pure power law form of the mass function, as opposed to the Schechter form. The corresponding likelihood function for the power law model is given by:
\begin{equation}
    p({M_i}| \theta, \tau) \equiv  \prod_{i=1}^{N} \frac{1}{Z_i} ({M_i})^{\alpha} p_{obs}(M_i| \tau),
\end{equation}
where the normalization becomes
\begin{equation}
    Z_i = \int_{M_{50}}^{\infty}  ({M_i})^{\alpha} p_{obs}(M_i| \tau, N_{MS}).
\end{equation}

\section{Results}\label{sec: results}

\subsection{Schechter Function Fitting Results}

\begin{figure*}
    \centering
    \includegraphics[width=8.29cm]{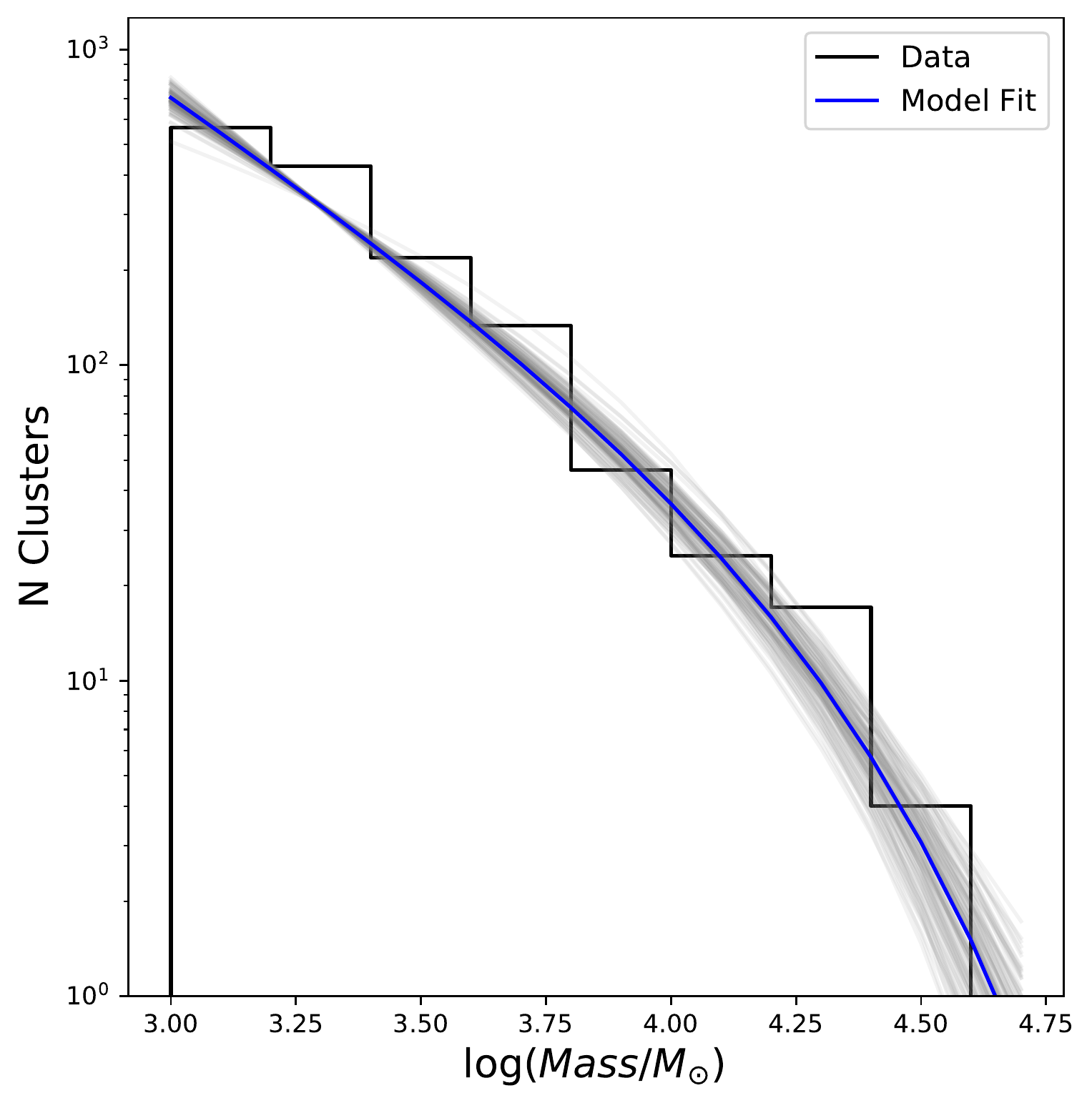}
    \includegraphics[width=8.71cm]{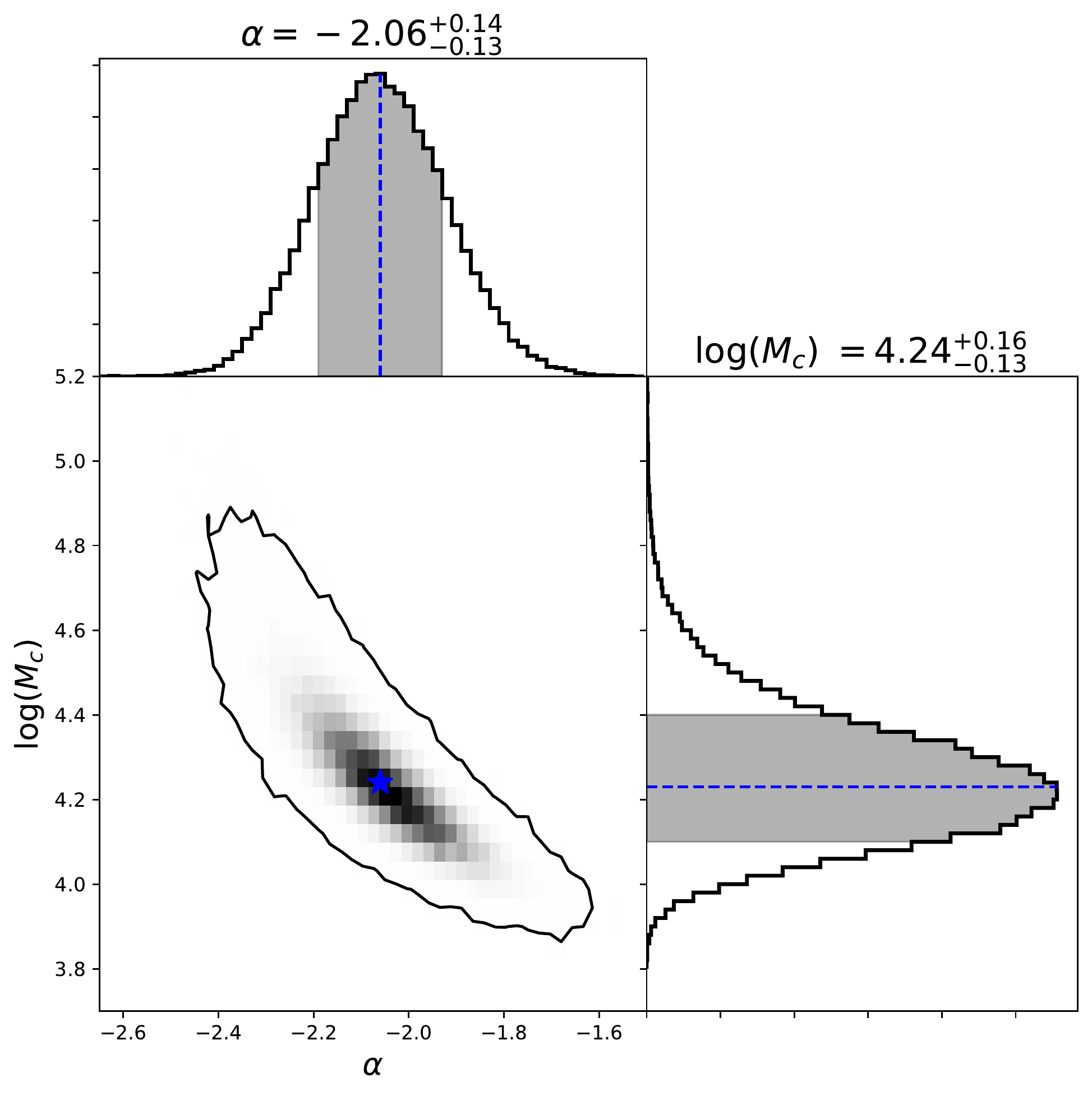}
    \caption{Schechter function fitting results for young star clusters in M33. {\em Left:} a histogram representing the completeness corrected mass distribution (black). The blue line visualizes the median posterior PDF values from Schechter function fits, and in gray we show 100 random samples from the posterior PDF to show the variance in the fits. We stress that the binned histogram is used only for visualization purposes and the fitting is performed on unbinned data. {\em Right:} the two dimensional posterior constraints on power law slope, $\alpha$, and Schechter truncation mass, $M_c$. The contour represents the $3\sigma$ limits of the two-dimensional PDF, taken as the 98.89 percentile of the density distribution. The blue star represents where the median values of the one-dimensional PDFs lie with respect to the two-dimensional density. The additional panels show the marginalized one-dimensional PDFs for Schechter parameters, with the blue dashed line showing the median of the distribution and the shaded in region the $1\sigma$ (16th and 84th percentile) confidence intervals. }
    \label{fig:main_result}
\end{figure*}

Schechter function fitting results for 533 young clusters with masses greater than the local 50\% mass completeness limit are shown in Figure \ref{fig:main_result}. The left panel shows the completeness corrected mass distribution for the observed sample and a Schechter function overplotted in blue that uses median $\alpha$ and $M_c$ values derived from marginalized posterior PDFs. We draw 100 random ($\alpha$, $M_c$) samples from the 2D posterior PDF and plot their corresponding Schechter forms in gray to indicate the variance in the fitted parameters. The binned histogram of observed masses is used only for visualization purposes, and not used for Schechter function fitting.

We find the CMF to be well described by a Schechter function with $\alpha = -2.06^{+0.14}_{-0.13}$, and log($M_c/M_{\odot}$) $= 4.24^{+0.16}_{-0.13}$. These results are based on marginalized one-dimensional posterior PDFs shown in the right panel of Figure~\ref{fig:main_result}.   We discuss this primary result further and place it into context with previous results in Section~\ref{sec: discusion}.  

\subsubsection{CMF Dependencies on Cluster Properties} \label{ssec: age_dep}

\begin{figure}
    \centering
    \includegraphics[scale=0.5]{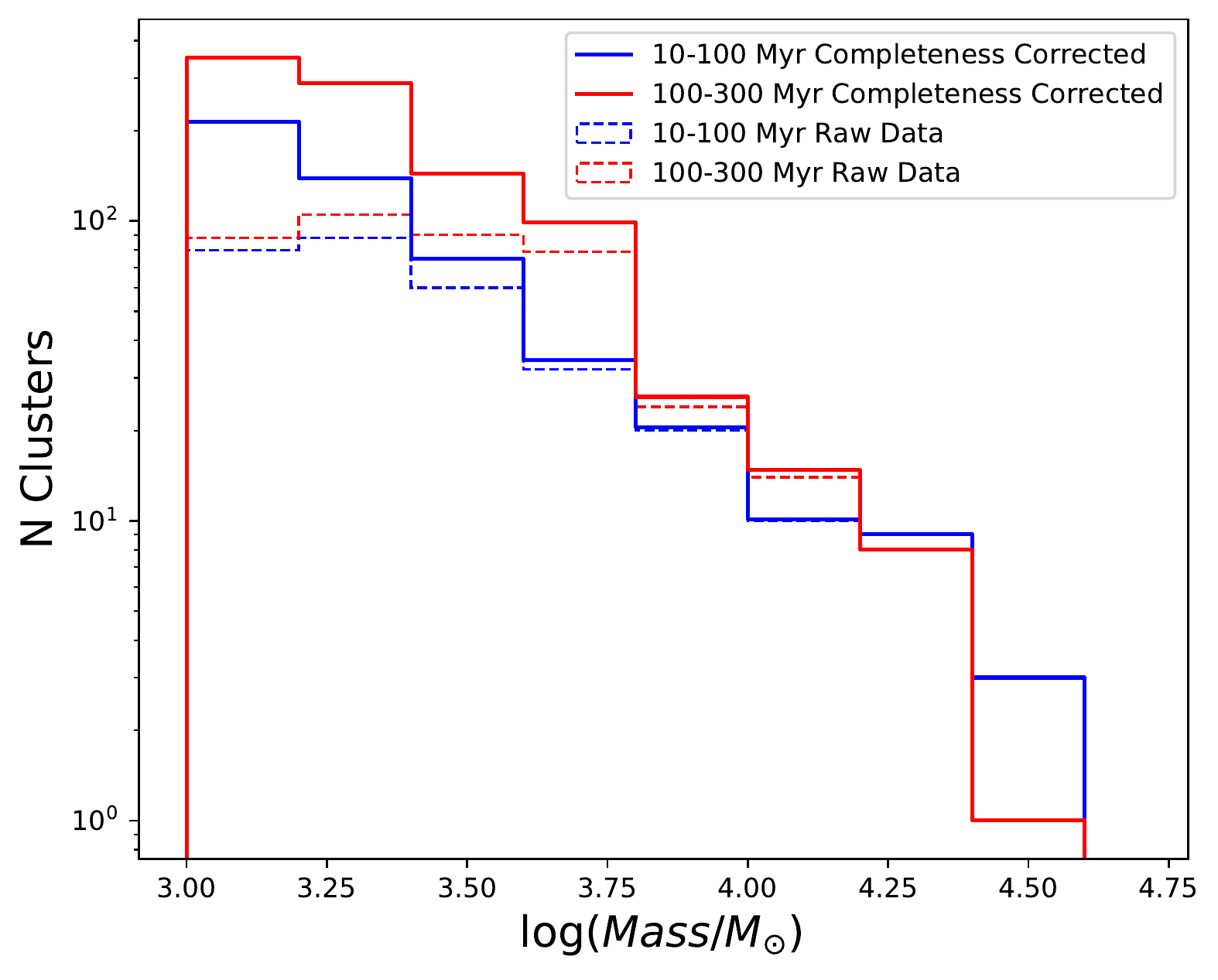}
    \includegraphics[scale=0.42]{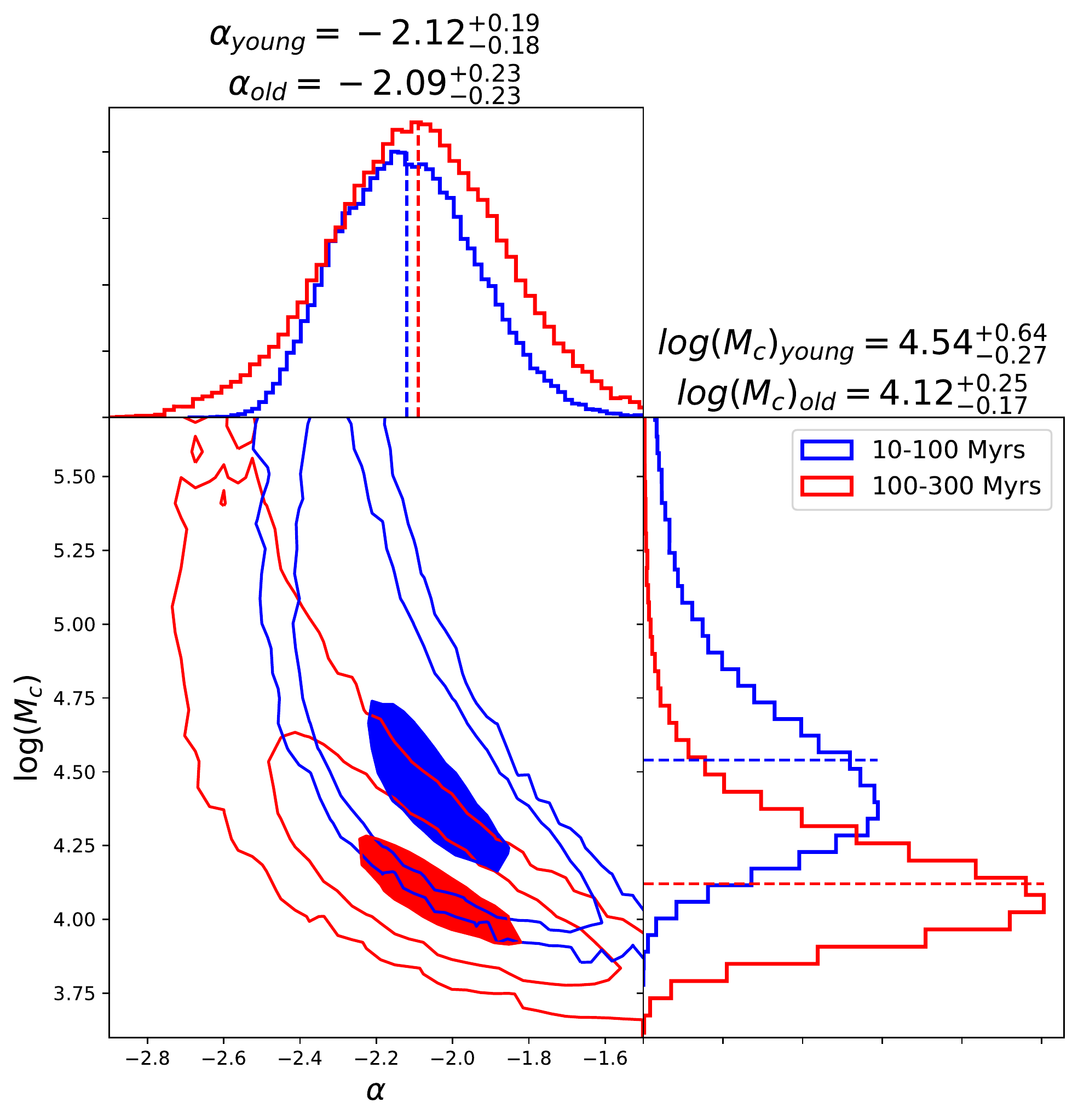}
    \caption{Mass function results split by age. {\em Top:} The observed mass distribution for the clusters split into two bins of cluster age. Blue represents clusters with ages between 10-100 Myrs and red represents clusters with ages between 100-300 Myrs. We present raw counts in the dashed lines and the completeness-corrected distribution in the solid lines. {\em Bottom:} the two dimensional posterior constraints on the Schechter function $\alpha$ and $M_c$ for each age bin. The contours represents the  $1\sigma, 2\sigma ,3\sigma$ range of the 2D density, where the $1\sigma$ range is shaded in. Additional panels show the marginalized one-dimensional PDFs for Schechter parameters, with the dashed line showing the median of the distribution for each age bin.}
    \label{fig:age_comp}
\end{figure}

We test for age dependence of the Schechter function fits by dividing the cluster sample into age bins of 10-100 Myrs and 100-300 Myrs, as shown in the top panel of Figure \ref{fig:age_comp}. A signature of mass dependent cluster destruction would be a flattening of the low mass slope of the CMF with increasing age due to the dissolution of low mass clusters \citep{gieles_what_2009}. Using clusters with masses greater than the local 50\% mass completeness, we fit 254 clusters with ages between 10-100 Myrs and 279 clusters with ages between 100-300 Myrs. We present Schechter function parameters for the two age bins in the bottom panel of Figure \ref{fig:age_comp}. 

In addition to the median values presented in Figure~\ref{fig:age_comp} we find the maximum a posteriori best-fit values to be, $\alpha = -2.04$, and $-2.03$, and log($M_c/M_{\odot}$) $=4.39$ and $4.07$ for the young and old samples respectively. Both values well within the 1$\sigma$ range of the PDF's.

We do not observe any flattening in the slope of the CMF with increased age. There is therefore no clear evidence for mass-dependent cluster destruction being important over time scales of 300 Myrs in the central regions of M33. We do notice an increase in $M_c$ for the younger sample. In comparing the PDFs, we find an 87$\%$ probability that the younger sample has a higher $M_c$ than the older sample. This could be evidence of a non-constant SFH, with enhanced star formation in the last 100 Myr relative to the 100-300~Myr time period. However, we note that the constraints on the best-fit Schechter function parameters are significantly broader than for the full sample due to the decrease in number statistics in each age bin. Therefore, drawing strong conclusions about M33's SFH based on these measurements is difficult.  

In addition to age, the CMF could also depend on galactic radius. Previous studies have shown a radial dependence on Schechter function truncation mass, where $M_c$ decreases with galactic radius \citep{adamo_probing_2015}, and in M33 there is evidence of a radial dependence on stellar age, with older ages at the center \citep{williams_detection_2009, davidge_stellar_2011}. Unfortunately, this work is unable to answer this question. The region of M33 imaged by the PHATTER survey only extends to galactic radii of $\sim$5~kpc, and we only find 92 clusters with radii $>$3 kpc. Fits to radially binned samples yielded no significant trends due to these limited number statistics and the small range of galactic radii probed.  Determination of any CMF radial dependence in M33 will require cluster samples that extend to larger radii.

\subsubsection{Analyzing our Assumptions in Completeness}

The parameters of the Schechter function fits are affected by our adopted completeness model. We examine the size of potential systematic errors based on our completeness corrections and sample selection. We do this by comparing our cluster-by-cluster completeness function to the simpler binned completeness function of \citet{johnson_panchromatic_2017}.

First, we directly use the methodology of \citet{johnson_panchromatic_2017}, binning the sample by age, and calculating the completeness on a binned basis opposed to a cluster-by-cluster basis. We split the sample into bins of 10-100~Myr and 100-300~Myrs. We find 589 clusters with masses greater than the age bins 50\% completeness, which leads to fitted Schechter function parameters $\alpha$ $=-1.90$, and log($M_c/M_{\odot}$) $= 4.25$. It is interesting to note that the power law slope for the binned correction is slightly outside of the 1$\sigma$ range of our fitted PDF, while the $M_c$ parameter is extremely similar. Further, this power law slope is similar to the value published for M31, suggesting that having M31's completeness calculated on a cluster-by-cluster basis would bring the same power law slope for M31 and M33 into agreement.

We also analyze the impact of our choice to include environmental dependence in our completeness function. To remove this dependence, we use Equation~\ref{eq: comp_from_cat_pap}, and assume the sample wide $c$ parameter fit in \citet{johnson_catalog_2022} to be 2.9770. After recalculating the local 50\% completeness limits  without the environmental dependence, we find 578 clusters have masses greater than the local 50\% completeness limit. Schechter function fitting results in the power law slope $\alpha$ flattened slightly to $-1.98$, with an log($M_c/M_{\odot}$) of 4.31. However, both values are well within the $1\sigma$ uncertainty of our fitted PDF.

To conclude the comparison of our completeness function to \citet{johnson_panchromatic_2017}, we believe our enhanced completeness function is justified. Even so, the results from a simplified completeness model seem to yield results within the 1$\sigma$ uncertainty. More importantly, there is no need to refit M31, and we believe we can reasonably compare the results from the two completeness models.

All of these models only include clusters above the local 50\% completeness limit.  To test how varying this minimum completeness impacts our results, we varied the minimum completeness from 40\% to 90\%.  We find for values $>$50\% (which we tested at 55\%, 60\%, 75\%, 80\%, and 90\%) that changes are within 1$\sigma$ of our results presented above.  However, at lower  completeness, the best-fit parameters started to deviate from our best-fit results.  Specifically, if we lower the completeness limit to 45\% the best fit parameters are a power law slope $\alpha = -2.30$ and truncation mass log($M_c/M_{\odot}$) $= 4.49$. If we lower it further to a 40\% completeness limit we find the best fit parameters are a power law slope $\alpha = -2.49$ and truncation mass log($M_c/M_{\odot}$) $= 4.68$.  In summary, using clusters above the local 50\% completeness limit appears to give robust results consistent with more conservative completeness limits, however, including clusters at lower completeness starts to impact our results significantly. 

\subsection{Power Law Fitting Results and Comparison to Schechter Function Fits}

\begin{figure*}[ht]
    \centering
    \includegraphics[width=6cm]{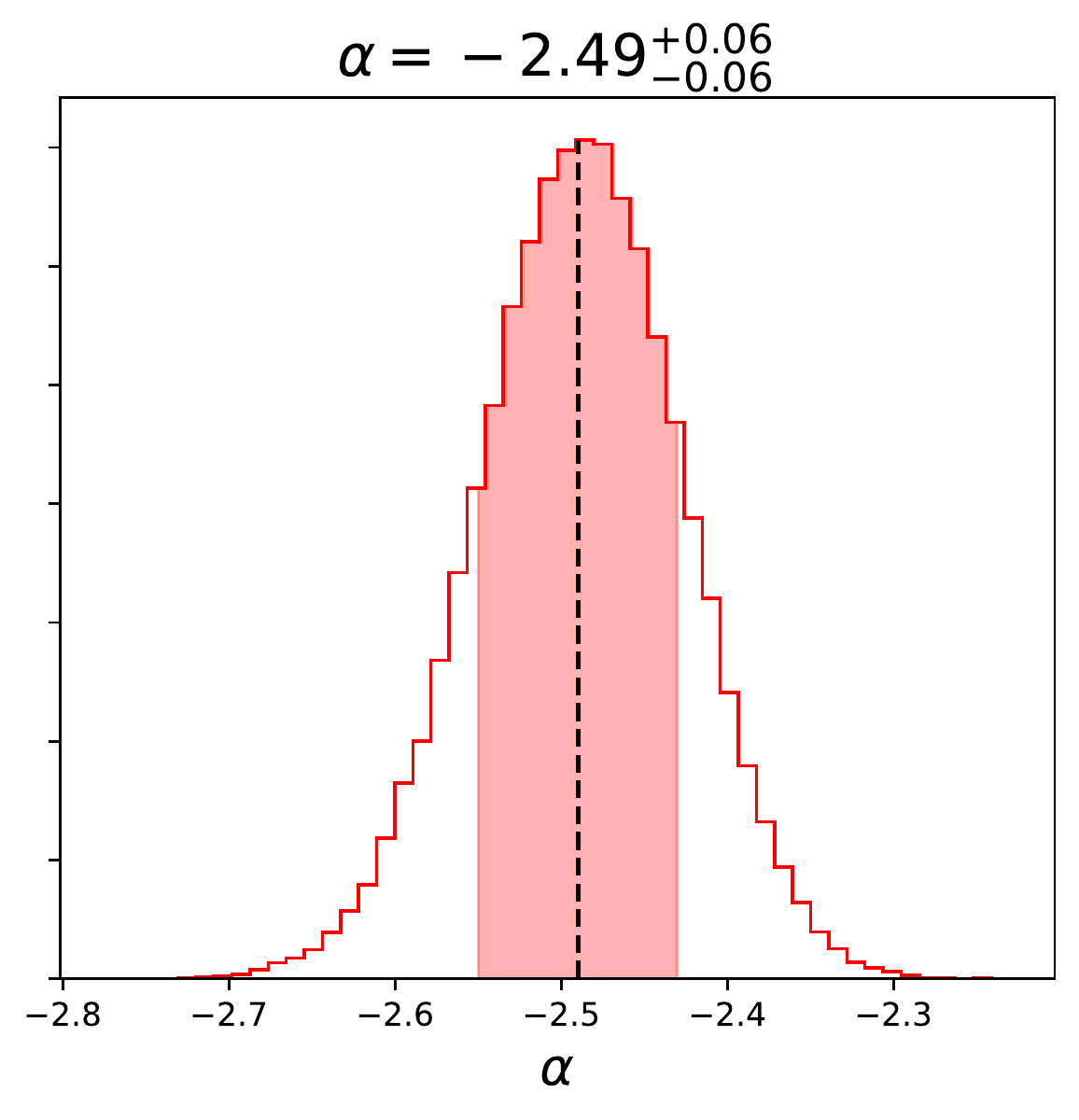}
    \includegraphics[width=5.35cm]{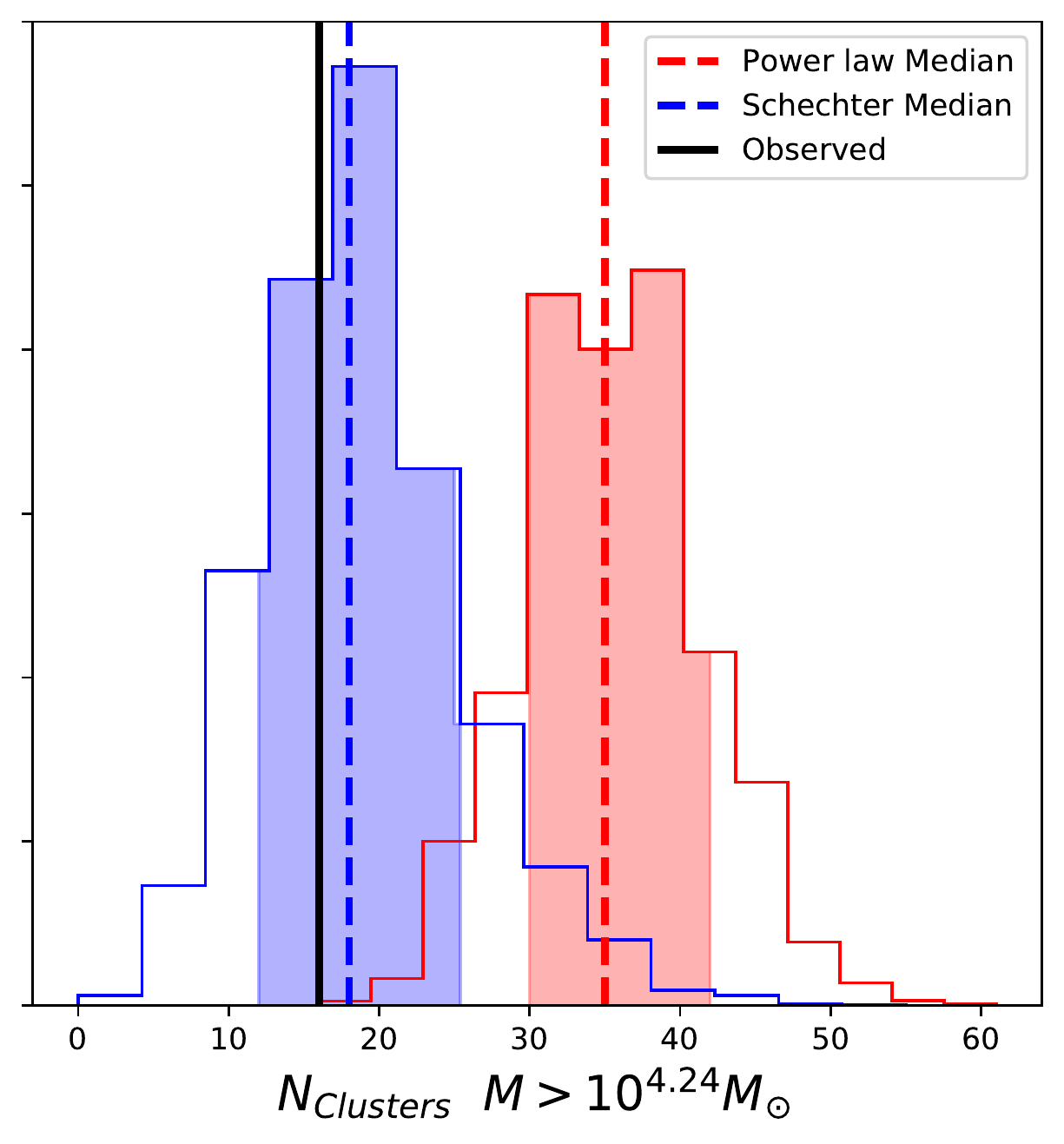}
    \includegraphics[width=6cm]{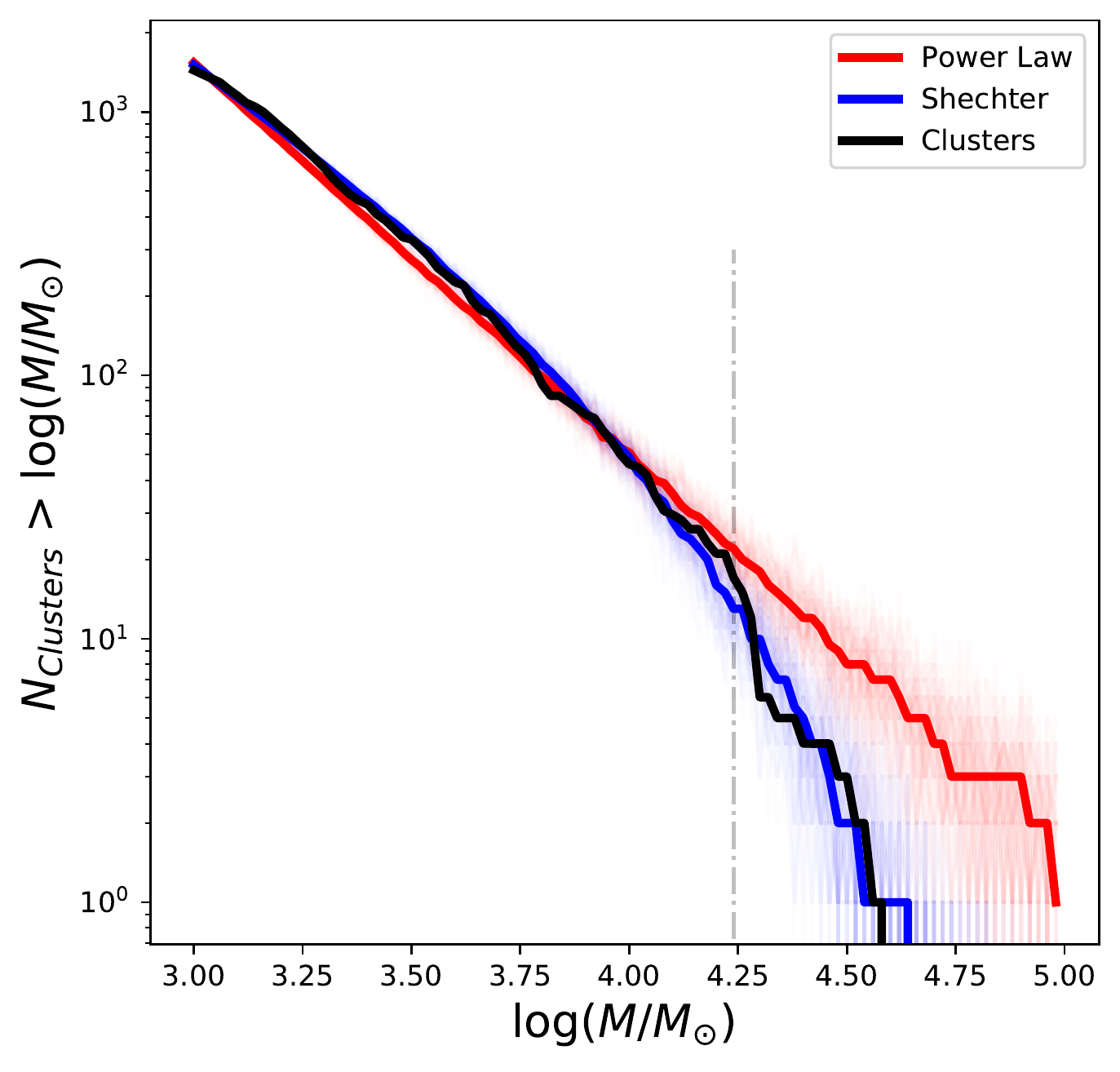}
    \caption{Pure power-law fit results and a comparison to Schechter function fits. {\em Left:} the marginalized PDF for power law fitting results where the black line represent the median of the distribution and the red shaded region are the $1\sigma$ uncertainties. {\em Middle:} the number of clusters above the derived $M_c$ value of $10^{4.24} M_{\odot}$. The black vertical shows the observed number of clusters above this mass. The Schechter (blue) and power law (red) model distributions were determined by sampling 10,000 draws from each PDF and counting the total number of high mass clusters above $10^{4.24} M_{\odot}$. The dashed lines represent the median values, and the shaded regions are the $1\sigma$ uncertainties to the nearest whole cluster. {\em Right:} the cumulative number of clusters above the mass given on the x-axis. Black represents the observed counts in M33, which we compare to number counts from the Schechter function (blue) and power law (red) models. For the models, we draw 100 random distributions with parameters of the median fitted PDF, and plot the median of the 100 samples in solid lines. }
    \label{fig:pl_plots}
\end{figure*}

In addition to fitting a Schechter function, we fit the M33 cluster sample for a power law distribution. Power law fitting results are shown in the left panel of Figure \ref{fig:pl_plots}. We find the distribution of clusters with masses greater than the local 50\% mass completeness limit to be best described by $\alpha = -2.49^{+0.06}_{-0.06}$. This slope is much steeper than the canonical -2 power law in the literature \citep[e.g.][]{krumholz_star_2019}, but still overpredicts the number of high mass clusters by $>4\sigma$.  

Previous studies have used the number of massive clusters to test whether cluster masses follow a power law or Schechter function distribution  \citep{johnson_panchromatic_2017, adamo_formation_2020}. Implementing this same test here, we observe 16 M33 clusters above the fitted truncation mass of log($M_c/M_{\odot}$) $= 4.24$. To compare the models to this observation, we calculate this same number for 10,000 synthetic distributions of clusters drawn from the PDFs of both the Schechter function and power law fits presented above. 

The predictions from these synthetic distributions are shown in the right two panels of Figure~\ref{fig:pl_plots}. In both panels it is evident that the Schechter function distribution better matches the actual number of high mass clusters (i.e. 16). The middle panel shows that the median number of synthetic clusters observed above log($M/M_{\odot}$) of $4.24$ for the Schechter function is $18^{+7}_{-6}$, and $35^{+7}_{-5}$ for the power law, where uncertainties give the 16th and 84th percentiles of the distributions. All 10,000 draws from the power law distribution result in more than the 16 observed high-mass clusters. The Schechter function provides a much better fit to this population of high-mass clusters.  Tests using other threshold masses indicate that these results are not sensitive to our exact threshold choice.

The right panel of Fig.~\ref{fig:pl_plots} shows the cumulative distribution of clusters above the mass on the x-axis.  We perform a Kolmogorov–Smirnov (KS) test on these distributions for both Schechter and power law distributions to assess the overall goodness of fit to observed data. Due to inherent randomness in drawing distributions, we draw 100 distributions and report the median values of cluster counts above a given mass to compare cumulative distributions to observed counts, as shown in the middle of Figure \ref{fig:pl_plots}. The KS statistic for the Schechter function is 0.04 with a p-value of 0.99 indicating the observed distribution is consistent with the Schechter function. However, for the power law distribution the KS statistic is 0.21 with a p-value of 0.02. Thus, the test suggests that the two samples are not drawn from the same distribution. In addition to the KS test, we run an Anderson-Darling test, which suggests a Schechter function is consistent with our mass distribution with  $p > 0.25$; on the other hand, the power law distribution has a p-value of $0.008$, and thus is not consistent with the data. 

One final test we use to determine which model best describes the cluster sample is the Bayesian Information Criterion (BIC) test. We find a delta BIC of 6.28. We use the \citet{kass_bayes_1995} criterion which classifies this delta BIC as strong evidence in favor of Schechter function parameters opposed to the power law model. We also perform this test on the two age samples in Section~\ref{ssec: age_dep}. Due to the decreased number statistics, the delta BIC is less significant than for the full sample at 5.54 and 5.64 for the young and old samples respectively, but each sample nonetheless provides positive evidence in favor of the Schechter function according to the \citet{kass_bayes_1995} guidelines. 

Both the number of high mass clusters, and mass distribution of clusters is therefore not well described by a power law distribution, thus providing strong evidence for the existence of a truncation at higher masses.  

\section{Discussion}\label{sec: discusion}

In this section we discuss the relation between $M_c$ and \SigSFR\ for M33 and compare it to other galaxies. We also discuss the implications of individual mass uncertainties on CMF results both in our, and previous literature results. 

\subsection{M33 CMF in Context: Comparison to Observational Results and Theoretical Predictions}

\citet{johnson_panchromatic_2017} found a clear correlation between the truncation mass of the CMF and a galaxy's SFR surface density, \SigSFR. In this section we assess whether M33 follows this trend, and compare our results to theoretical predictions of mass function truncation.

Following the methodology described in Appendix A of \citet{johnson_panchromatic_2017}, we measure a characteristic \SigSFR\ for M33. We combine \textit{GALEX} FUV and \textit{Spitzer} 24 \textit{$\mu$m} images following the prescription of \citet{leroy_star_2008} to produce a map of \SigSFR. We use this particular SFR prescription for consistency with other galaxy measurements, but note that in the future a CMD-based star formation history estimate will be available for the PHATTER survey region (M. Lazzarini, in prep.). We use an SFR-weighted average surface density, $\langle \Sigma_{SFR} \rangle$, to summarize the local, kpc-scale properties of M33's disk in a way that accounts for the non-uniform distribution of star formation. We measure log($\langle \Sigma_{SFR} \rangle$~/~(\Msun~yr$^{-1}$~kpc$^{-2})$) of $-2.04^{+0.16}_{-0.18}$ for M33 within the PHATTER survey footprint.

\begin{figure}
    \centering
    \includegraphics[width=8.5cm]{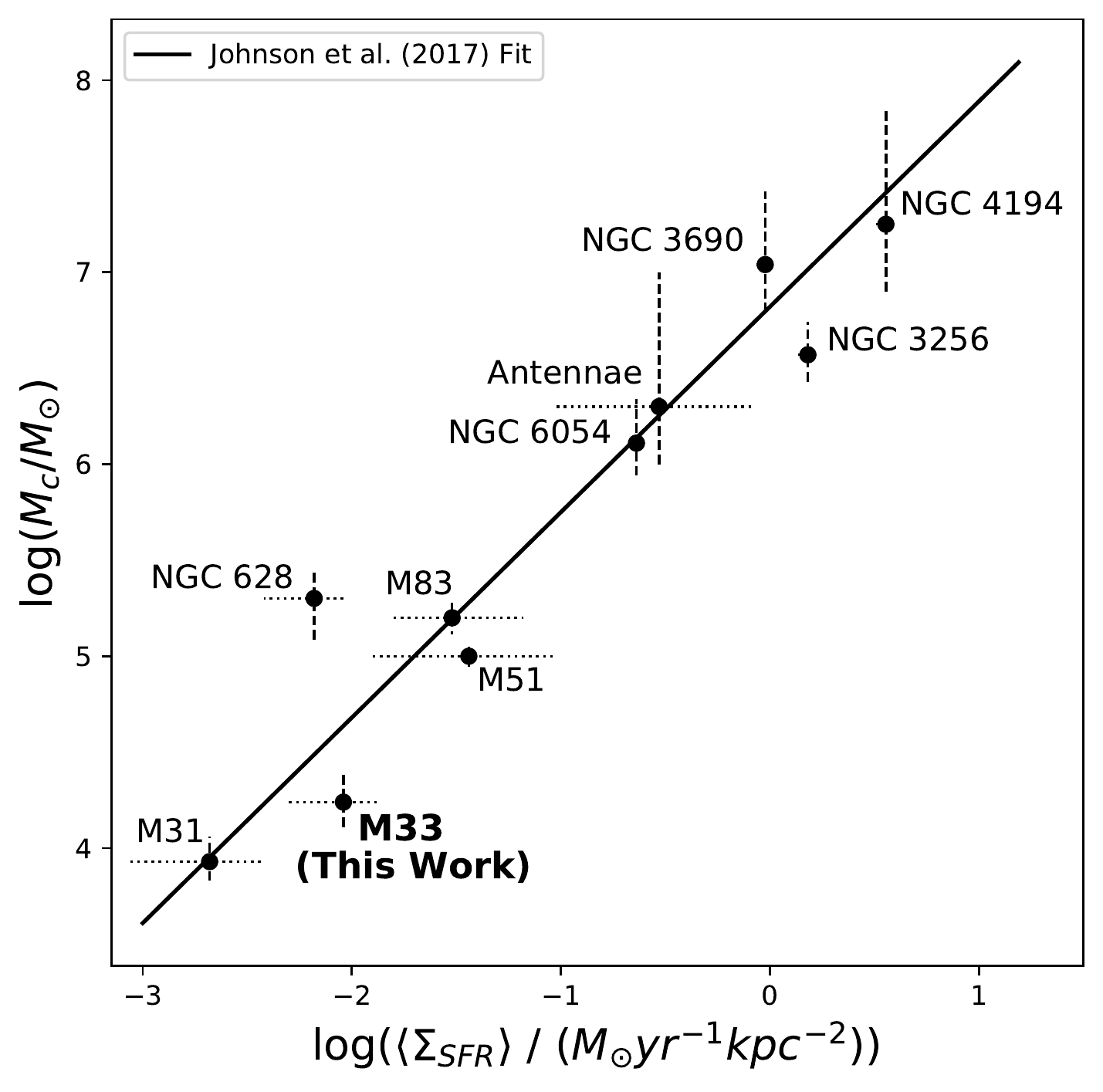}
    \caption{Comparison of log($M_c$) values as a function of average SFR surface density $\langle \Sigma_{SFR} \rangle$ for young star clusters.  The following data are included: M31 \citep{johnson_panchromatic_2017}, M33 (this work), M83 \citep{adamo_probing_2015}, NGC628 \citep{adamo_legacy_2017}, M51 \citep{messa_young_2018}, the Antennae galaxies \citep{zhang_mass_1999, jordan_acs_2007}, NGC 3256, NGC 3690, NGC 4194, and NGC 6054 \citep{adamo_formation_2020}. Dashed vertical lines denote uncertainties in $M_c$, while the dotted horizontal lines represent the narrowest 68\% interpercentile range for local \SigSFR\ measurements. The black line represents the fit of \citet{johnson_panchromatic_2017}, described in Equation~\ref{eq:MC_vs_SSFR}.}
    \label{fig:MC_vs_SSFR}
\end{figure}

We compare $M_c$ and $\langle \Sigma_{SFR} \rangle$ for M33 to a compilation of measurements for other galaxies in Figure \ref{fig:MC_vs_SSFR}. We combine results from M31 \citep{johnson_panchromatic_2017}, M51 \citep{messa_young_2018}, M83 \citep{adamo_probing_2015}, NGC628 \citep{adamo_legacy_2017}, the Antennae \citep{jordan_acs_2007}, as well as four galaxies from the HiPEEC survey \citep[NGC 3256, NGC 3690, NGC 4194, and NGC 6054;][]{adamo_formation_2020}. For $\langle \Sigma_{SFR} \rangle$ measurements, we adopt the 80\% \SigSFR\ values tabulated in \citet{adamo_formation_2020} for the HiPEEC galaxies, while we use measurements from \citet{johnson_panchromatic_2017} for M31, M51, M83 and the Antennae.  While the HiPEEC 80\% \SigSFR\ values are not a perfect match to the SFR-weighted $\langle \Sigma_{SFR} \rangle$ measurements discussed above, this alternative form of area-weighting tends to produce a similar, focused measure of \SigSFR.

\citet{johnson_panchromatic_2017} found a nearly linear relation between $M_c$ and $\langle \Sigma_{SFR} \rangle$ represented by, 
\begin{equation} \label{eq:MC_vs_SSFR}
    \log(M_c\mathrm{)} = (1.07 \pm 0.10) \times \log( \langle \Sigma_{SFR} \rangle \mathrm{)} + (6.82 \pm 0.20).
\end{equation}
We plot this relation as the black line in Figure~\ref{fig:MC_vs_SSFR}. In \citet{johnson_panchromatic_2017}, the fit was based on just four data points: M31, M51\footnote{Note that the M51 $M_c$ measurement used here was updated to the result from \citet{messa_young_2018} as opposed to \citet{gieles_what_2009}.}, M83, and the Antennae.  Here, we add six new data points that show remarkable agreement with the original fitted trend from Equation~\ref{eq:MC_vs_SSFR}, with a mean and maximum residual of 0.24 dex and 0.81 dex, respectively. We also use the Python package \texttt{linmix} \citep{kelly_aspects_2007} to fit this new sample of observations and constrain the relation's intrinsic scatter, accounting for uncertainties in $M_c$ and \CSigSFR. We derive a slope of $0.97 \pm 0.13$ and an intercept of $6.56 \pm 0.18$, which are consistent with the values fit by \citet{johnson_panchromatic_2017}. We constrain a median intrinsic scatter of only $0.19^{+0.39}_{-0.15}$ dex around the newly fit relation, which is small relative to the four dex range of both axes. 

In addition to the observational results discussed above, \citet{reina-campos_unified_2017} present a theoretical model that extends the cluster formation efficiency ($\Gamma$) prescription from \citet{kruijssen_fraction_2012} and predicts a maximum cluster mass based on three environmental observables: the gas surface density, the epicylic frequency, and the Toomre $Q$ parameter. We use rotation curve and radial surface density profiles of total gas and stars compiled in \citet{utomo_origin_2019} to compute a prediction for the PHATTER M33 cluster sample.  We adopt input parameter values\footnote{\SigGas\ = 11.0 \solperpc, $\Omega$ = 0.0367 Myr$^{-1}$, Toomre Q = 4.21, and $\phi_P$ = 9.51 \citep[used as part of $\Gamma$ calculation; see \S 4.2 in][]{johnson_panchromatic_2016}.} that correspond to a characteristic galactic radius of 2.2 kpc, the median galactocentric radius of the PHATTER cluster sample.

The \citet{reina-campos_unified_2017} model predicts a feedback-limited maximum cluster mass for M33 of log($M/M_{\odot}$) = 4.06, $\sim$0.2 dex below our fitted $M_c$ value.  Similarly, we calculate a model prediction for M31 using observable input values\footnote{\SigGas\ = 10.5 \solperpc, $\Omega$ = 0.021 Myr$^{-1}$, Toomre Q = 1.77, and $\phi_P$ = 1.6.} applicable to the dominant 10 kpc "Ring-Total" region from the PHAT $\Gamma$ study by \citet{johnson_panchromatic_2016}, whose properties apply to a vast majority of that sample's clusters. The \citet{reina-campos_unified_2017} model predicts a feedback-limited maximum cluster mass for M31 of log($M/M_{\odot}$) = 3.53, 0.4 dex below the fitted $M_c$ value.  While the model appears to underestimate the maximum cluster mass in both M31 and M33, the availability of an environmentally-dependent model that is accurate within a factor of 2-3 is a fantastic step toward better understanding cluster formation. 

The nearest galaxy to M33 in $\langle \Sigma_{SFR} \rangle$ is NGC628; however there is $\sim 1$ dex difference in values of log($M_c/M_{\odot}$). Despite the similar $\langle \Sigma_{SFR} \rangle$, these two galaxies are quite different, and these differences may account for their different $M_c$ values. Specifically, NGC~628 has a log($M_\star$) of 10.2 \citep{cook_spitzer_2014}, nearly an order of magnitude higher mass than M33.  Similarly, NGC~628's peak rotation curve velocity is $\sim$180~km/s \citep{aniyan_resolving_2018}, about 60\% higher than in M33 \citep{utomo_origin_2019}. Also, NGC~628's central metallicity is  [O/H]$=$+0.1 \citep{berg_chaos_2015}, roughly 0.2 dex higher than M33.  Further, as we will discuss in Section~\ref{sec: discusion}, the large mass errors on the individual cluster measurements could lead to a significant upward bias in the measurement of $M_c$ in NGC~628 that may account for some of the observed disagreement with M33.  

\subsection{Effect of Individual Cluster Mass Uncertainties on the CMF}

Our CMF fitting method ignores the individual cluster mass errors that results from CMD analysis. For small errors ($\lesssim$0.05~dex), this simplification has been shown to minimally impact the mass function fitting \citep{weisz_panchromatic_2013}. However, larger errors may result in significant biases in our inferred parameters. In particular, integrated light measurements typically have larger mass errors than the CMD based mass estimates we have used here. In this section, we explore the potential biases in mass function parameters as a function of the cluster mass error distribution by using literature samples of both CMD and integrated light measurements. We expect large mass errors to lead to the overestimation of $M_c$ values, due to the steeply declining mass function, which results in more clusters being scattered to higher mass values.  

\subsubsection{M31 Results: Impacts on the CMF from Integrated Light Masses} \label{ssc: M31_IL}

\begin{figure}
    \centering
    \includegraphics[width=8.5cm]{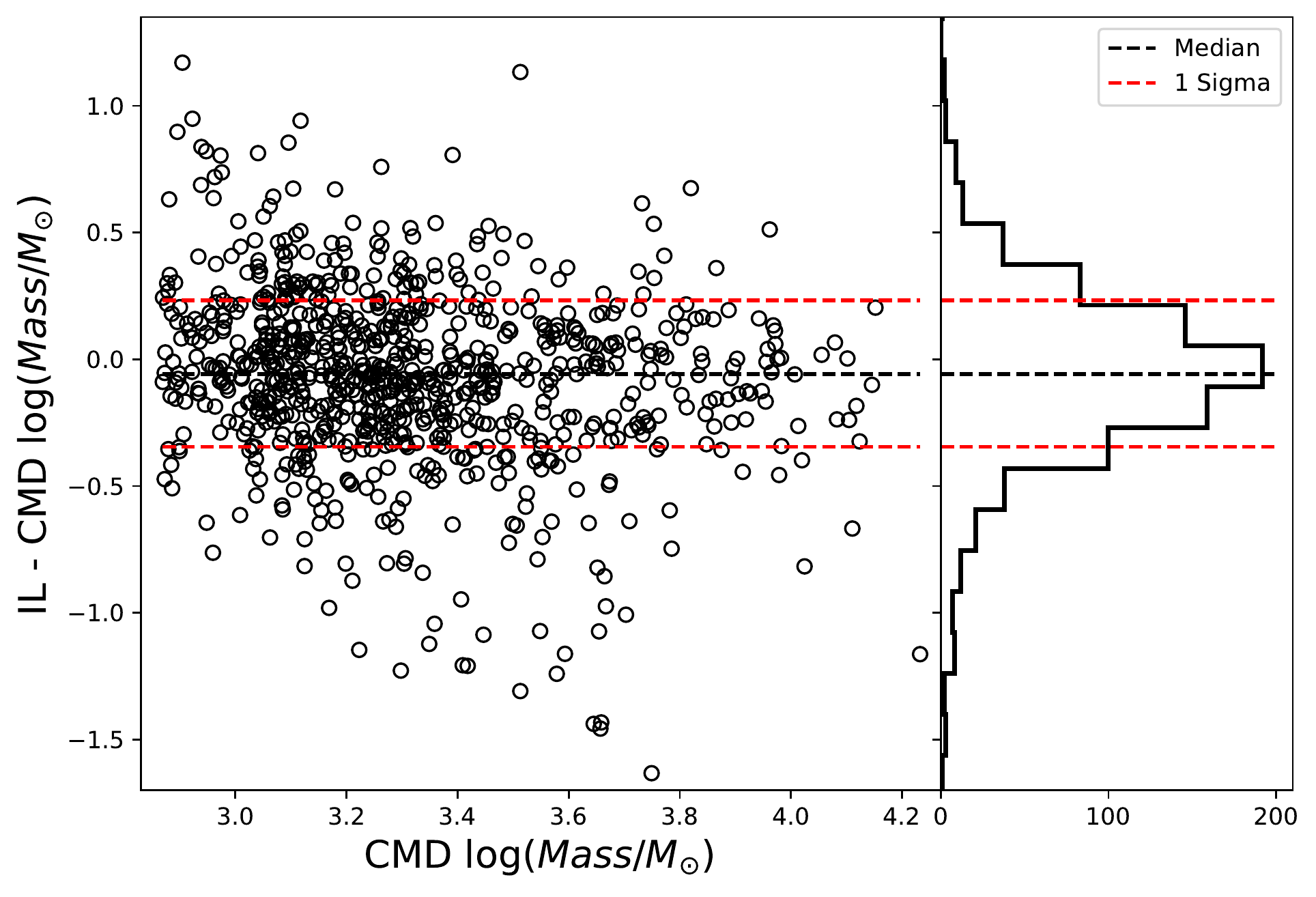}
    \caption{The comparison between M31 cluster masses derived from CMD analysis of \citet{johnson_panchromatic_2017} and integrated light (IL) analysis of M. Fouesneau (\textit{pri. comm.}). The y-axis shows the difference between the integrated light and CMD mass estimates, plotted against the CMD mass estimate. Clusters are represented by the open black circles, where the dashed black line represents the median of the distribution, and the red dashed lines represent the 1$\sigma$ range. The Additional panel shows the one dimensional distribution of the differences. }
    \label{fig:M31_mass_comp}
\end{figure}

We first consider data from M31 where cluster masses have been derived with both CMD and integrated light (IL) methods. The CMD masses in M31 \citep{johnson_panchromatic_2016} were derived using an identical technique to that described in Section~\ref{sec: data}. The cluster ages used range between 10 and 300 Myr. The integrated light ages and masses were fit by M. Fouesneau (\textit{pri. comm.}) using the method described in \citet{fouesneau_panchromatic_2014}. The authors utilize a large number of simulated clusters that are sampled using MCMC to develop a PDF for each cluster's age, mass and extinction. We show the direct comparison of CMD masses to the integrated light masses in Figure~\ref{fig:M31_mass_comp}. Because we use the same sample of clusters that have CMD ages, for our mass function fitting we use the same completeness function of \citet{johnson_panchromatic_2017} for both cluster samples.

\begin{figure}
    \centering
    \includegraphics[width=8.5cm]{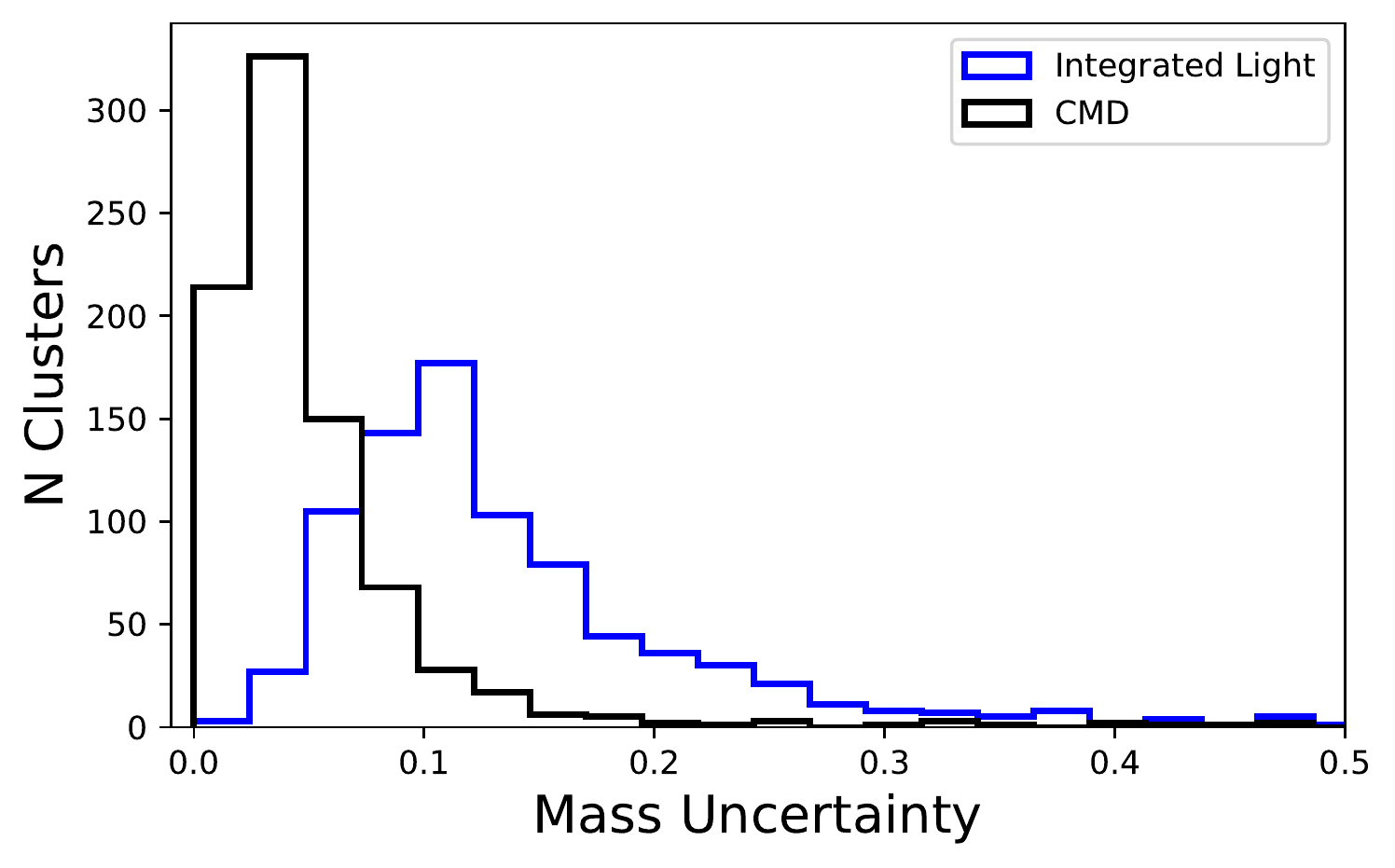}
    \caption{Cluster mass error distributions for M31 clusters. In black are are mass errors from \citet{johnson_panchromatic_2017} derived through CMD analysis, while the blue line shows the same clusters with properties derived through SED integrated light fitting (M. Fouesneau \textit{priv. comm.}).}
    \label{fig:M31_errs}
\end{figure}

We quantify the distributions of reported uncertainties for the two methods in Figure~\ref{fig:M31_errs}. The median 1$\sigma$ errors on the cluster masses from the CMD method are just 0.03 dex.  However, for the integrated light measurements the median 1$\sigma$ errors are 0.14 dex. Further, the median ratio of integrated light uncertainty to CMD uncertainty is 3.3. While the difference in uncertainties between the two methods is large, in Figure~\ref{fig:M31_mass_comp}, we do not see a significant bias in the one-to-one comparison of cluster masses, with the median difference of $-0.06$ dex. However, there is a fair amount of scatter around this median with a standard deviation of $0.59$ dex.

\begin{figure}
    \centering
    \includegraphics[width=8.5cm]{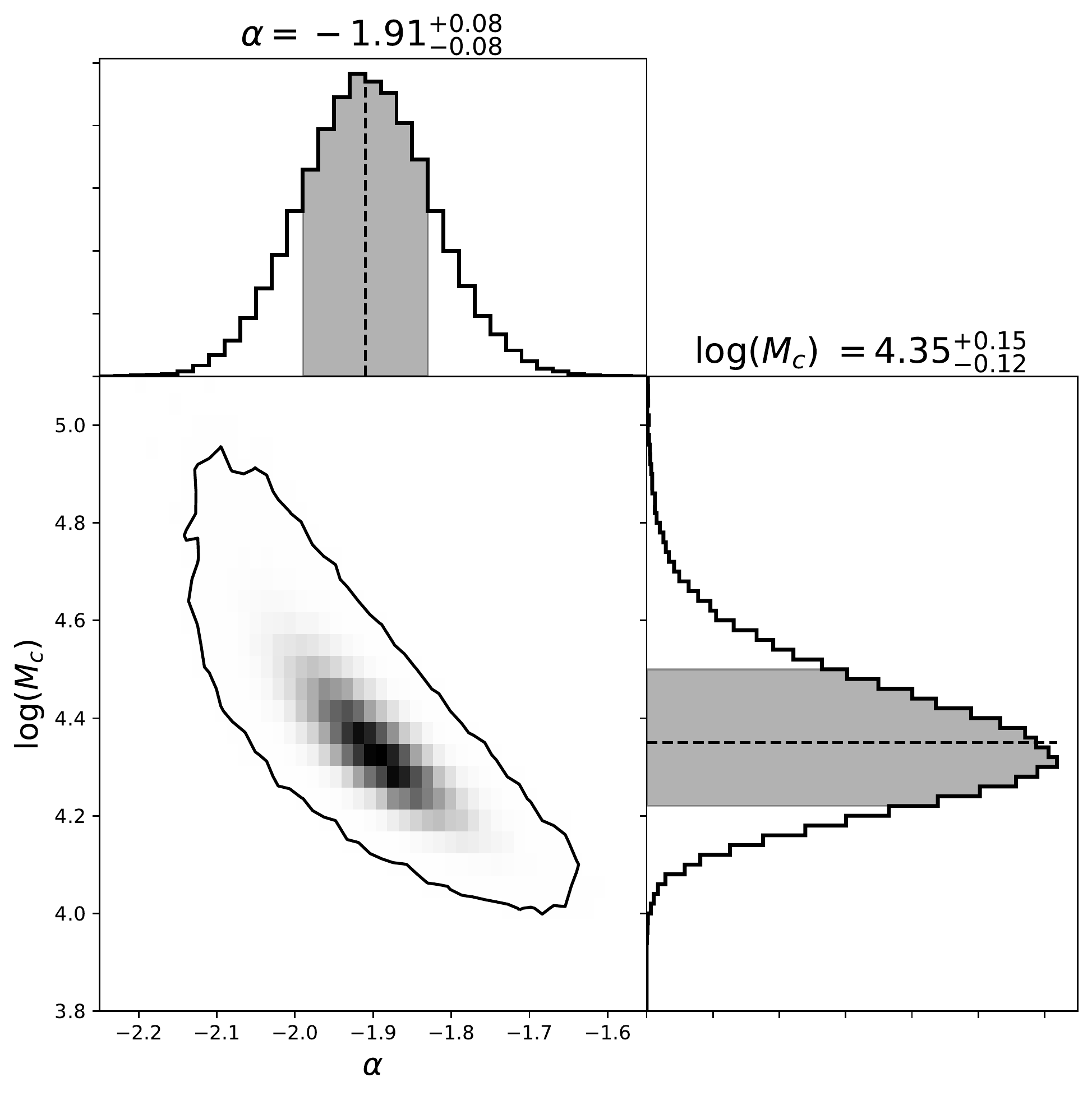}
    \caption{Schechter function fitting results for M31 clusters using age and mass estimates determined from integrated light fitting (M. Fouesneau \textit{priv. comm.}). We present the two-dimensional PDF of Schechter function fitting results. The contour represents the $3\sigma$ limits of the two-dimensional PDF, taken as the 98.89 percentile of the density distribution. Additional panels show the marginalized one-dimensional PDFs for Schechter parameters, with the black dashed line showing the median, and shaded region the $1\sigma$ uncertainty range.}
    \label{fig:M31_IL}
\end{figure}

The Schechter function fits to the CMD based masses is presented in \citet{johnson_panchromatic_2017}. We use an identical method to fit the integrated light ages for the same sample of clusters; the fit results are shown in in Figure \ref{fig:M31_IL}. The best Schechter function has a power law index of $\alpha = -1.91 \pm 0.08$ and a truncation mass of log$(M_c/\odot)$ = $4.35^{+0.15}_{-0.12}$. For the CMD masses, \cite{johnson_panchromatic_2017} finds $\alpha = -1.99 \pm 0.12$, with log($M_{c}/M_{\odot}$) $= 3.93^{+0.13}_{-0.10}$. 

Based on the analysis for M31, we find that the inferred value of $M_{c}$ is 0.4 dex higher when masses are measured using the less accurate integrated light method rather than the higher accuracy CMD masses. This result is very striking. An 0.4 dex increase in $M_c$ translates to a factor of $\sim2.5$ in cluster mass, and biases of this level could significantly impact the trend of $M_c$ with \SigSFR. It is important to note that the only difference between the two measurements is the method in which the ages and masses were derived. 

\subsubsection{Examining $M_c$ Biases in Published Mass Function Fits}
To test if cluster mass uncertainties are the cause of the observed bias in the previous section, in this section we run a series of tests fitting mock samples of clusters drawn from a Schechter function, incorporating the effects of mass uncertainty. In our first experiment we examine the effect of a fixed mass error for each cluster, in the second, we insert single high mass outliers, while the third experiment incorporates the observed error distributions and Schechter function parameters of galaxies with published CMF fits. Overall, we find that large individual cluster mass uncertainties leads to a significant positive bias for inferred $M_c$ values.  

We supplement clusters from M31 with data from M51, M83 and NGC628. The integrated light cluster mass data from these catalogs are from the M83 results in \citet{adamo_probing_2015}, and the LEGUS results \citep{calzetti_legacy_2015}: \citet{adamo_legacy_2017} for NGC628, and \citet{messa_young_2018} for M51.  A detailed description of the cluster identification and mass measurements is given in \citet{adamo_legacy_2017}. 
Due to the distance of these galaxies ($\sim4$ Mpc, $\sim$5$\times$ that of M31 and M33), clusters appear only partially resolved, and thus only integrated light measurements are possible. We use the Bayesian mass estimates of the clusters derived using the SLUG code \citep{krumholz_slug_2015}.  Their sample includes clusters with masses above 5000~M$_\odot$. 

In all cases, we assume the mass uncertainty estimates to be Gaussian in log($M$), where the standard deviation of the Gaussian correspond to the ($84th$ percentile - $16th$ percentile) $ / 2$ of the PDF. We note that while we refer to these errors as mass errors, they are actually errors in log($M/M_{\odot}$). We incorporate these individual cluster mass errors into simulated mass function fits.

For our first experiment, we create a synthetic distribution of 1000 clusters with $\alpha$, and log($M_{c}/M_{\odot}$) values of $-2$, and $5.0$ respectively, consistent with the best fit parameters of typical literature galaxies. We fit the initial distribution as a baseline, then add a uniform Gaussian mass error to each cluster. We then refit a Schechter function to the new distribution of scattered masses. Finally, we compare the fitted $M_{c}$ parameter to the value of the original distribution. We run 100 trials of this experiment for a fixed mass error of 0.05 dex, and 0.15 dex, which covers the range of error distributions we observe in the literature. 

For a mass error of 0.05 dex, we find a median log($M_c/M_{\odot}$) bias of 0.01 dex above the original value, while for the 0.15 dex mass errors we get a log($M_c/M_{\odot}$) bias of 0.09 dex. When we compare these biases to the median 1$\sigma$ range for the derived log($M_c/M_{\odot}$) parameters of 0.30 dex, these biases are not outside of normal uncertainty. This base experiment shows that for small errors, the impact on Schechter function parameters is minimal, but that larger errors produce more bias in the Schechter function truncation masses.

\begin{figure}
    \centering
    \includegraphics[width=8.5cm]{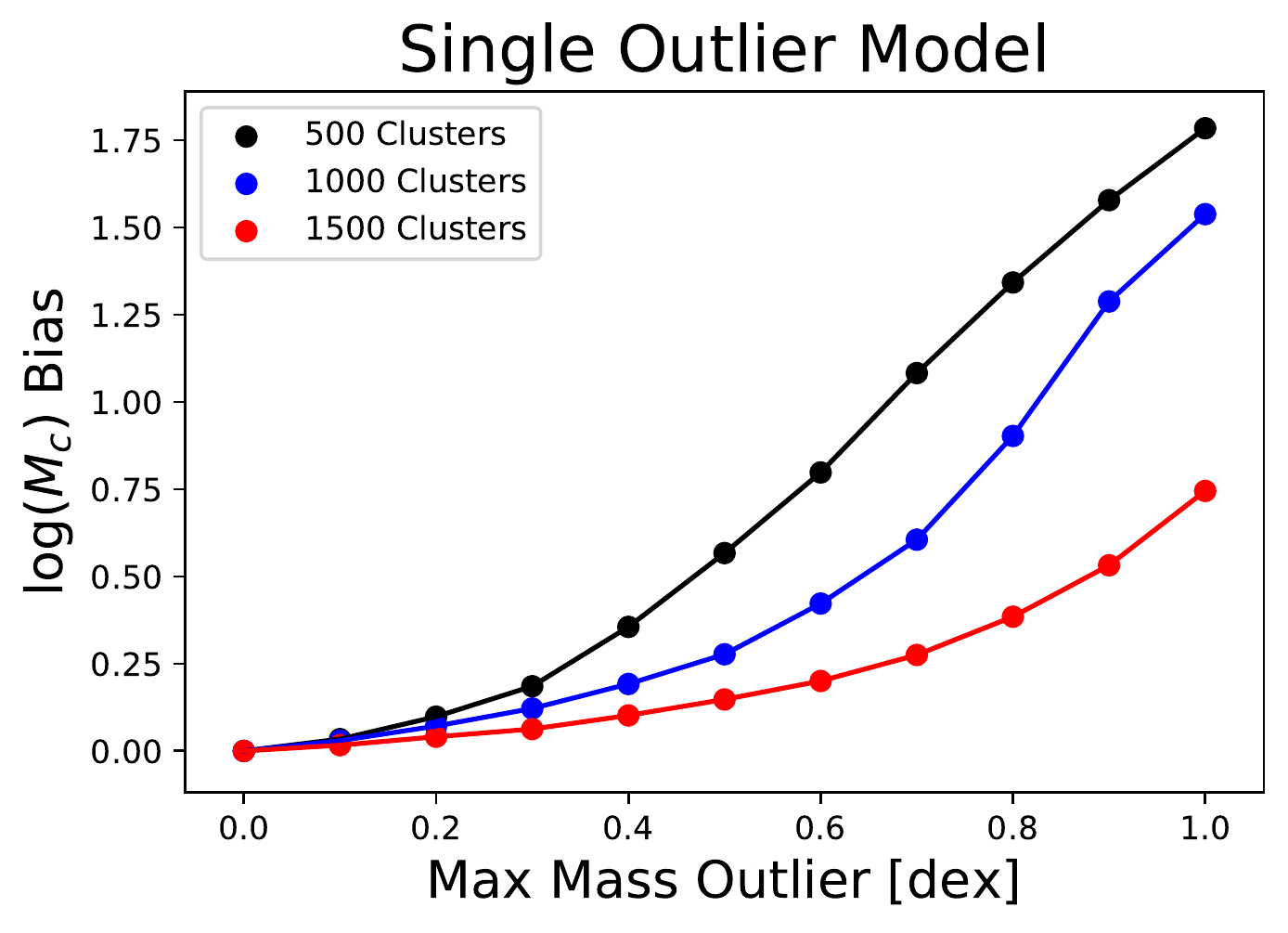}
    \caption{The bias in the Schechter mass function truncation $M_c$ due to high mass outliers. Points represent what happens to the estimated $M_c$ for cluster mass distribution drawn from a Schechter function, but with a single cluster moved to a higher mass. The x-axis indicates the difference in the highest mass cluster between the original and the altered distributions. The y-axis shows the difference in derived log($M_c$) from the input distribution. The different colors represent different numbers of clusters in the drawn distributions. The larger the mass outlier, the larger the resulting bias in $M_c$.}
    \label{fig:Walking}
\end{figure}

During this first experiment we found a handful of our simulated clusters were scattered to very high masses. To understand the effect that a single high mass outlier has on the best-fit CMF, we conduct a second experiment. We use the same setup as in our first experiment with no mass errors added. We then add a single, high mass cluster starting from the maximum mass cluster drawn from the Schechter function (which had log($M/M_{\odot}$) = 5.35), and steadily increasing this mass to log($M/M_{\odot}$)=6.35. The results are shown in Figure~\ref{fig:Walking}. As expected, the single high mass cluster raises the $M_c$ value inferred, with very significant biases seen for maximum mass outliers more than 1 dex higher than the rest of the distribution. We run this experiment for a range of input $M_c$ values and total number of clusters that reflects values of the galaxies we discuss here (i.e M31, M33, M51, M83, NGC628) and find these results to be consistent across input parameters. This shows the significant impact cluster mass errors can have for the highest mass clusters. While a single high mass cluster can impact the inferred $M_c$ value, the majority of the observed bias is not due to single outliers.

For each of the three samples presented in Figure~\ref{fig:Walking}, we calculate the delta BIC for Schechter parameters relative to the pure power law model. Not surprisingly, we find that with increased number statistics, the delta BIC favoring a Schechter model is slightly larger. However, regardless of number statistics, the single outlier causes negligible changes to the delta BIC, and in all cases the delta BIC is larger than 6, providing strong evidence for a Schechter function model over a power law model according to the \citet{kass_bayes_1995} guidelines. While the Schechter function is still preferred, for large bias, the inferred $M_c$ values are very poorly constrained with 1$\sigma$ uncertainties upwards of $\sim$1 dex. These constraints are worse for the sample with the fewest clusters.

Our third experiment simulates the real distribution of uncertainty in published mass estimates by generating a synthetic cluster distributions from Schechter function parameters published in the literature and assigning each synthetic cluster a mass error of an associated real cluster. We then add a Gaussian random mass error to each cluster, and fit a Schechter function to the new distribution of scattered masses running an MCMC fit. Finally, we compare the fitted $M_{c}$ parameter to the value fitted to the original distribution. We run 100 trials per galaxy. We note that we are able to assign a real cluster mass error to a random synthetic cluster because we do not find a correlation between cluster mass and mass error in any of the cluster samples considered here.

We run this experiment for M83 and NGC628 using the Schechter function fits from \cite{adamo_probing_2015, adamo_legacy_2017} respectively, while for M51 we use the results in \cite{messa_young_2018}. We also run the experiment on our set of CMD derived cluster properties in M31 \citep{johnson_panchromatic_2017}, and M33 (this work), as well as the SED integrated light cluster properties for M31 clusters discussed in Section~\ref{ssc: M31_IL}. For each galaxy we report the median output log($M_c$) - input log($M_c$) for the 100 trials, along with the $1\sigma$ confidence interval. 

\begin{figure}
    \centering
    \includegraphics[width=8.5cm]{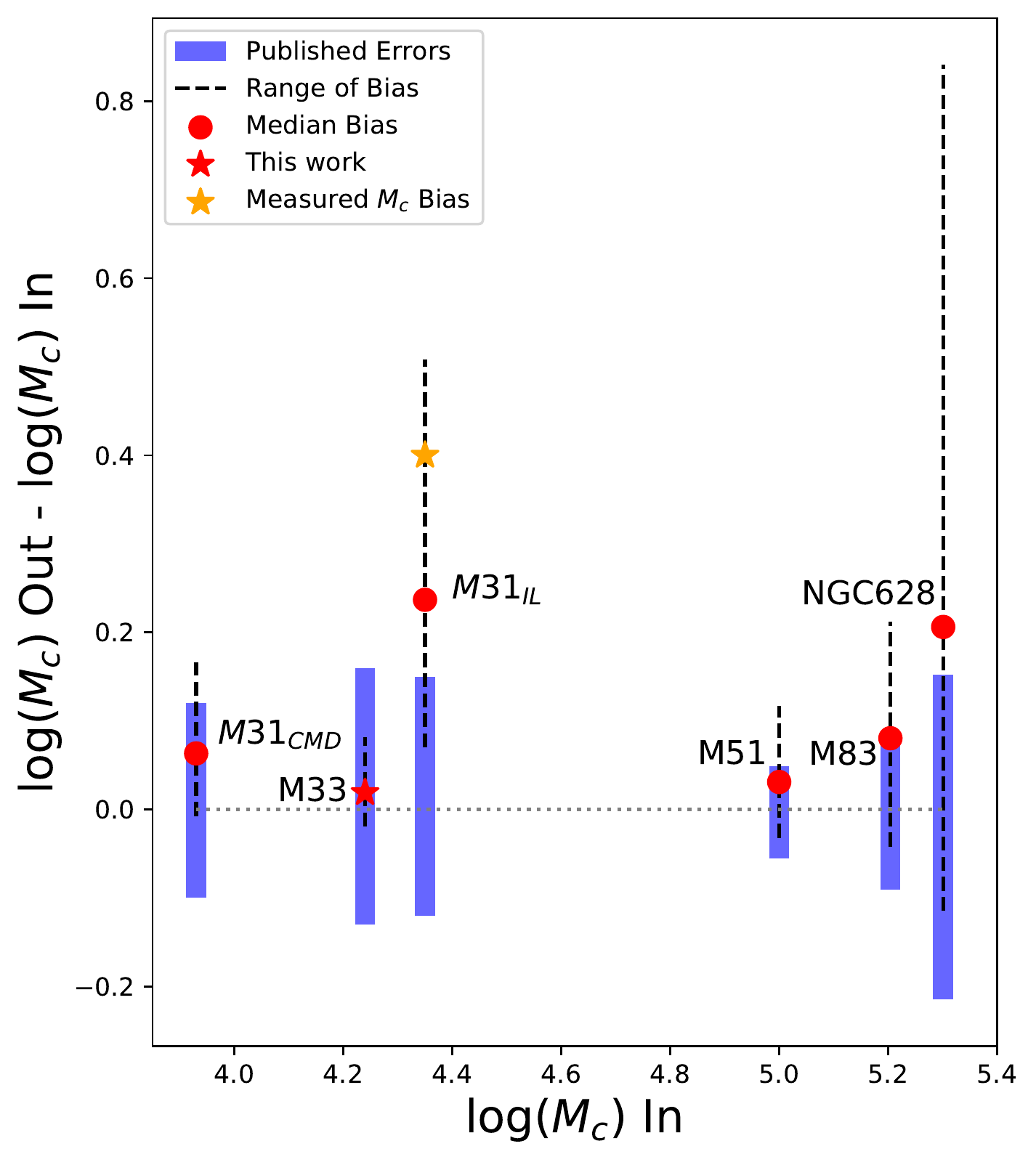}
    \caption{The bias in Schechter function truncation mass $M_c$ for observed data sets.  The y-axis indicates the difference in log($M_c$) values between simulations without (input) and with (output) mass errors. Red points indicate the median differences of 100 trials, while black dashed lines represent the 1 $\sigma$ range of results. The blue solid lines are the published confidence intervals centered around the dashed gray line at 0. We present experiment results for M33 (this work), M31 (CMD \citep{johnson_panchromatic_2017}; integrated light (M. Fouesneau \textit{priv. comm.})), M51 \citep{messa_young_2018}, M83 \citep{adamo_probing_2015}, and NGC628 \citep{adamo_legacy_2017}. The orange star represents the measured delta $M_c$ between the CMD and integrated light measurements in M31 and shows that the observed difference falls within the distribution of the simulated differences.}
    \label{fig:Other_gals}
\end{figure}

We present the results of this experiment in Figure~\ref{fig:Other_gals}. We find that when we simulate realistic cluster mass uncertainties and fit the CMF, the $M_c$ value recovered is typically higher than the input value. The resulting $M_c$ bias is $<$1~dex in all cases, with the median bias being comparable to the reported 1$\sigma$ uncertainties in $M_c$ in all cases. The median bias is highest in NGC~628 and for the integrated light measurements of M31 due to these samples larger cluster mass errors. NGC~628 also has the smallest number of clusters, while the galaxy with the largest number of clusters (M51) shows a very small bias. We also find that the measured bias in log($M_c$) for M31 clusters discussed in the previous subsection falls within the distribution of simulated differences.

We have seen that both large average cluster mass errors and single mass outliers can significantly impact our estimate of $M_c$. Thus, the distribution of cluster mass errors, not just the average error, is important. To quantify the error distributions of our literature cluster samples, we perform an Anderson-Darling test and find that these distributions can each be described by a log-normal distribution. A log-normal distribution is parameterized by a mean, $\mu$, and standard deviation, $\sigma$, as represented by:

\begin{equation}
    \frac{1}{x \sigma \sqrt{2 \pi}} exp(-\frac{(ln(x)-\mu)^2}{2\sigma^2})
\end{equation}.

\begin{figure}
    \centering
    \includegraphics[width=8.5cm]{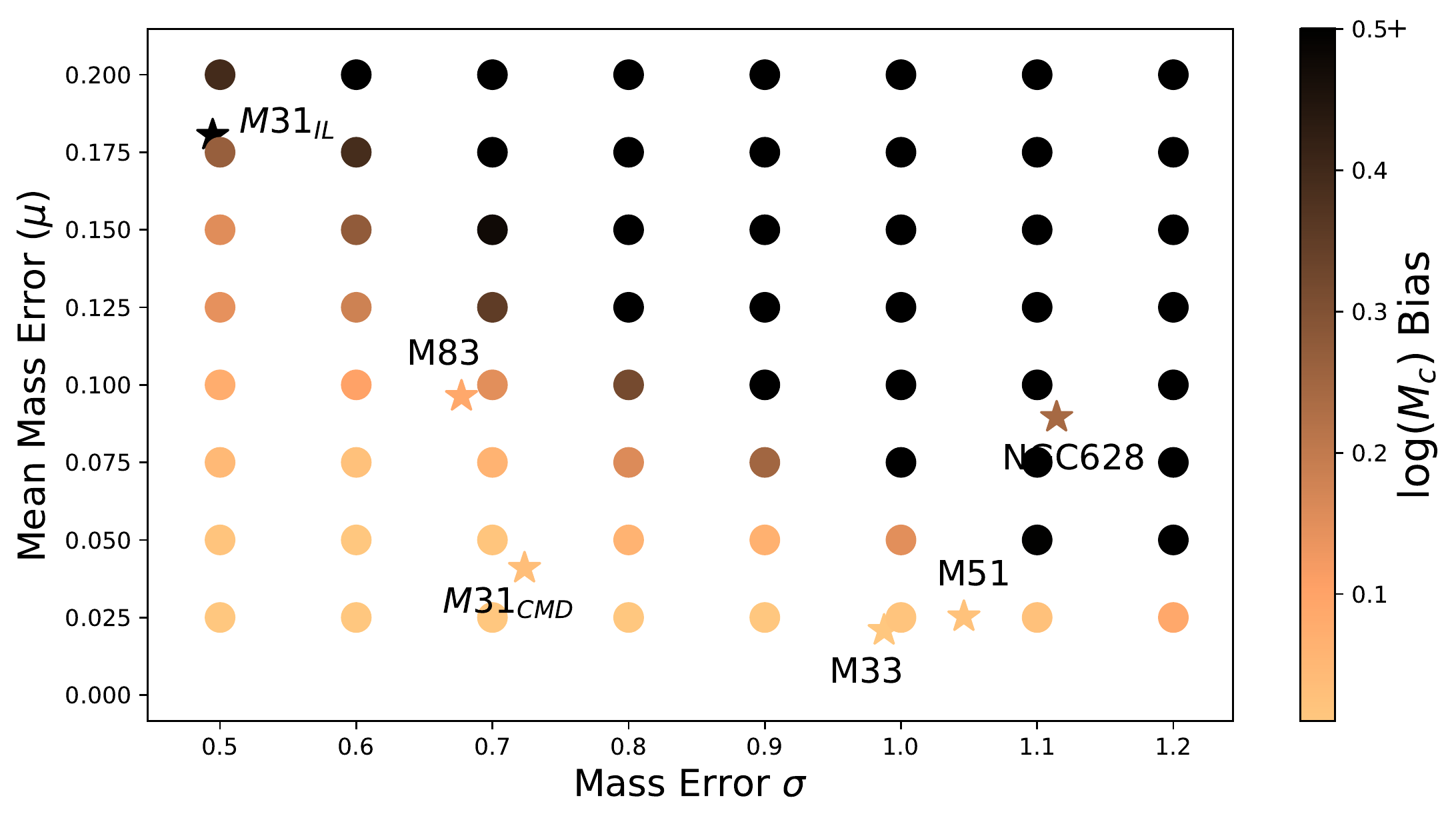}
    \caption{Bias in log($M_c$) estimates as a function of the distribution of individual cluster mass errors.  The two axes represent the mean error and width of a log-normal distribution of errors.  The grid of models show the resulting median $M_c$ bias from 100 trials with mass errors included. The stars represent the error distributions for each of the galaxies in Figure \ref{fig:Other_gals}, colored by the median bias found in Figure \ref{fig:Other_gals}.}
    \label{fig:Model_bias}
\end{figure}

For a log-normal's given $\mu$, $\sigma$, we calculate the predicted $M_c$ bias, shown in Figure~\ref{fig:Model_bias}. As in our first experiment, we assume 1000 cluster masses drawn from Schechter function parameters $\alpha = -2$ and $M_c = 5.0$ and fit this distribution as a baseline. Here we add cluster mass errors drawn from each log-normal distribution before refitting a Schechter function. We then take the median bias in $M_c$ from 100 trials.  These values are the circular points in Figure~\ref{fig:Model_bias}.

We find the model to have a clear gradient of increased bias with respect to both log-normal parameters. Thus, either having overall larger errors or having a tail of clusters with large errors can lead to significant biases in the inferred $M_c$. We also find the model to be consistent with the biases we observe in published fits.

However, we note that our test grid does not exactly match the number statistics of the individual measurements, which could lead to the small differences we see between the inferred values for the individual galaxies and our model grid. We verified that the calculated bias is insensitive to the input $M_c$.  

In practice, the $M_c$ bias could be avoided by incorporating cluster mass uncertainties into the derivation of the CMF with a hierarchical Bayesian model. For large cluster samples with small uncertainties (like for the M33 CMD mass estimates presented here) this more complicated procedure is not necessary. However, it should be considered for smaller samples, or those that rely on integrated light measurements.

\section{Summary} \label{sec: Summary}

We have used the star cluster catalog of \citet{johnson_catalog_2022} to measure the star cluster mass function within the PHATTER survey region of M33.
We find strong evidence for a high-mass truncation, where the data is best represented by Schechter function parameters with power law slope $\alpha = -2.06^{+0.14}_{-0.13}$, and truncation mass log($M_c/M_{\odot}$) $= 4.24^{+0.16}_{-0.13}$. We also show that the M33 CMF is not well described by a pure power law.

We derive a $\langle \Sigma_{SFR} \rangle$ value of $-2.04^{+0.16}_{-0.18}$ for M33. When we examine the relation between $M_c$ and \SigSFR, we find M33 agrees well the relation described in \citet{johnson_panchromatic_2017}, where $M_c \propto \langle \Sigma_{SFR} \rangle^{\sim1}$. Adding additional literature estimates for a total of 10 galaxies, we find the average residual from this relation is just 0.24 dex.  We fit the intrinsic scatter of the relation, and find it to be $0.19^{+0.39}_{-0.15}$ dex. This data adds evidence that there is a clear correlation between the cluster truncation mass $M_c$ and \SigSFR.

Finally, we analyse the effect individual cluster mass uncertainties have on mass function measurements. We find that the truncation mass can be biased to higher values for cluster samples with high mass uncertainties.   These biases are similar to the 1$\sigma$ errors on $M_c$ published in the literature.  We believe the next clear step in deriving reliable mass function fits is to develop a method for incorporating hierarchical Bayesian modeling into mass function derivations.
\vspace{5mm}


\acknowledgments

We recognize and thank the $\sim$2,800 Local Group Cluster Search volunteers who made this work possible. Their contributions are acknowledged individually at \url{http://authors.clustersearch.org}. Support for this work was provided by NASA through grant number HST-GO-14610 from the Space Telescope Science Institute, which is operated by AURA, Inc., under NASA contract NAS5-26555.  This material is partially based on work by T.M.W. as a Center for Interdisciplinary Exploration and Research in Astrophysics (CIERA) REU student at Northwestern University, supported by the National Science Foundation under grant No. AST-1757792. Additional support was provided from the Undergraduate Research Opportunities Program at the University of Utah awarded to T.M.W.. L.C.J. acknowledges support through a CIERA Postdoctoral Fellowship at Northwestern University. This work used computing resources provided by CIERA and Northwestern University. This research made use of NASA's Astrophysics Data System (ADS) bibliographic services.

\facilities{HST(ACS, WFC3)}
\software{emcee \citep{emcee2013}, corner \citep{corner2016}}


\bibliographystyle{aasjournal}
\bibliography{references}

\begin{thebibliography}{}
\expandafter\ifx\csname natexlab\endcsname\relax\def\natexlab#1{#1}\fi
\providecommand{\url}[1]{\href{#1}{#1}}

\bibitem[{Adamo {et~al.}(2015)Adamo, Kruijssen, Bastian, Silva-Villa, \&
  Ryon}]{adamo_probing_2015}
Adamo, A., Kruijssen, J. M.~D., Bastian, N., Silva-Villa, E., \& Ryon, J. 2015,
  Monthly Notices of the Royal Astronomical Society, 452, 246.
\newblock \url{http://adsabs.harvard.edu/abs/2015MNRAS.452..246A}

\bibitem[{Adamo {et~al.}(2017)Adamo, Ryon, Messa, Kim, Grasha, Cook, Calzetti,
  Lee, Whitmore, Elmegreen, Ubeda, Smith, Bright, Runnholm, Andrews, Fumagalli,
  Gouliermis, Kahre, Nair, Thilker, Walterbos, Wofford, Aloisi, Ashworth,
  Brown, Chandar, Christian, Cignoni, Clayton, Dale, de~Mink, Dobbs, Elmegreen,
  Evans, Gallagher, Grebel, Herrero, Hunter, Johnson, Kennicutt, Krumholz,
  Lennon, Levay, Martin, Nota, {\"O}stlin, Pellerin, Prieto, Regan, Sabbi,
  Sacchi, Schaerer, Schiminovich, Shabani, Tosi, Van~Dyk, \&
  Zackrisson}]{adamo_legacy_2017}
Adamo, A., Ryon, J.~E., Messa, M., {et~al.} 2017, The Astrophysical Journal,
  841, 131.
\newblock \url{http://adsabs.harvard.edu/abs/2017ApJ...841..131A}

\bibitem[{Adamo {et~al.}(2020)Adamo, Hollyhead, Messa, Ryon, Bajaj, Runnholm,
  Aalto, Calzetti, Gallagher, Hayes, Kruijssen, K{\"o}nig, Larsen, Melinder,
  Sabbi, Smith, \& {\"O}stlin}]{adamo_formation_2020}
Adamo, A., Hollyhead, K., Messa, M., {et~al.} 2020, Monthly Notices of the
  Royal Astronomical Society, 499, 3267, aDS Bibcode: 2020MNRAS.499.3267A.
\newblock \url{https://ui.adsabs.harvard.edu/abs/2020MNRAS.499.3267A}

\bibitem[{Allison {et~al.}(2010)Allison, Goodwin, Parker, Portegies~Zwart, \&
  de~Grijs}]{allison_early_2010}
Allison, R.~J., Goodwin, S.~P., Parker, R.~J., Portegies~Zwart, S.~F., \&
  de~Grijs, R. 2010, Monthly Notices of the Royal Astronomical Society, 407,
  1098, publisher: Oxford Academic.
\newblock \url{https://academic.oup.com/mnras/article/407/2/1098/1122061}

\bibitem[{Aniyan {et~al.}(2018)Aniyan, Freeman, Arnaboldi, Gerhard, Coccato,
  Fabricius, Kuijken, Merrifield, \& Ponomareva}]{aniyan_resolving_2018}
Aniyan, S., Freeman, K.~C., Arnaboldi, M., {et~al.} 2018, Monthly Notices of
  the Royal Astronomical Society, 476, 1909, aDS Bibcode: 2018MNRAS.476.1909A.
\newblock \url{https://ui.adsabs.harvard.edu/abs/2018MNRAS.476.1909A}

\bibitem[{Bastian {et~al.}(2012)Bastian, Adamo, Gieles, Silva-Villa, Lamers,
  Larsen, Smith, Konstantopoulos, \& Zackrisson}]{bastian_stellar_2012}
Bastian, N., Adamo, A., Gieles, M., {et~al.} 2012, Monthly Notices of the Royal
  Astronomical Society, 419, 2606.
\newblock \url{http://adsabs.harvard.edu/abs/2012MNRAS.419.2606B}

\bibitem[{Berg {et~al.}(2015)Berg, Skillman, Croxall, Pogge, Moustakas, \&
  Johnson-Groh}]{berg_chaos_2015}
Berg, D.~A., Skillman, E.~D., Croxall, K.~V., {et~al.} 2015, The Astrophysical
  Journal, 806, 16, aDS Bibcode: 2015ApJ...806...16B.
\newblock \url{https://ui.adsabs.harvard.edu/abs/2015ApJ...806...16B}

\bibitem[{Bik {et~al.}(2003)Bik, Lamers, Bastian, Panagia, \&
  Romaniello}]{bik_clusters_2003}
Bik, A., Lamers, H. J. G. L.~M., Bastian, N., Panagia, N., \& Romaniello, M.
  2003, Astronomy and Astrophysics, 397, 473.
\newblock \url{http://adsabs.harvard.edu/abs/2003A%26A...397..473B}

\bibitem[{Calzetti {et~al.}(2015)Calzetti, Lee, Sabbi, Adamo, Smith, Andrews,
  Ubeda, Bright, Thilker, Aloisi, Brown, Chandar, Christian, Cignoni, Clayton,
  da~Silva, de~Mink, Dobbs, Elmegreen, Elmegreen, Evans, Fumagalli, Gallagher,
  Gouliermis, Grebel, Herrero, Hunter, Johnson, Kennicutt, Kim, Krumholz,
  Lennon, Levay, Martin, Nair, Nota, {\"O}stlin, Pellerin, Prieto, Regan, Ryon,
  Schaerer, Schiminovich, Tosi, Van~Dyk, Walterbos, Whitmore, \&
  Wofford}]{calzetti_legacy_2015}
Calzetti, D., Lee, J.~C., Sabbi, E., {et~al.} 2015, The Astronomical Journal,
  149, 51.
\newblock \url{http://adsabs.harvard.edu/abs/2015AJ....149...51C}

\bibitem[{Chandar {et~al.}(2016)Chandar, Whitmore, Dinino, Kennicutt, Chien,
  Schinnerer, \& Meidt}]{chandar_age_2016}
Chandar, R., Whitmore, B.~C., Dinino, D., {et~al.} 2016, The Astrophysical
  Journal, 824, 71.
\newblock \url{http://adsabs.harvard.edu/abs/2016ApJ...824...71C}

\bibitem[{Cook {et~al.}(2014)Cook, Dale, Johnson, Van~Zee, Lee, Kennicutt,
  Calzetti, Staudaher, \& Engelbracht}]{cook_spitzer_2014}
Cook, D.~O., Dale, D.~A., Johnson, B.~D., {et~al.} 2014, Monthly Notices of the
  Royal Astronomical Society, 445, 899, aDS Bibcode: 2014MNRAS.445..899C.
\newblock \url{https://ui.adsabs.harvard.edu/abs/2014MNRAS.445..899C}

\bibitem[{Cook {et~al.}(2019)Cook, Lee, Adamo, Kim, Chandar, Whitmore, Mok,
  Ryon, Dale, Calzetti, Andrews, Aloisi, Ashworth, Bright, Brown, Christian,
  Cignoni, Clayton, da~Silva, de~Mink, Dobbs, Elmegreen, Elmegreen, Evans,
  Fumagalli, Gallagher~III, Gouliermis, Grasha, Grebel, Herrero, Hunter,
  Jensen, Johnson, Kahre, Kennicutt, Krumholz, Lee, Lennon, Linden, Martin,
  Messa, Nair, Nota, Ostlin, Parziale, Pellerin, Regan, Sabbi, Sacchi,
  Schaerer, Schiminovich, Shabani, Slane, Small, Smith, Smith, Taibi, Thilker,
  de~la Torre, Tosi, Turner, Ubeda, Van~Dyk, Walterbos, \&
  Wofford}]{cook_star_2019}
Cook, D.~O., Lee, J.~C., Adamo, A., {et~al.} 2019, Monthly Notices of the Royal
  Astronomical Society, 484, 4897, arXiv: 1902.00082.
\newblock \url{https://ui.adsabs.harvard.edu/abs/2019MNRAS.484.4897C/abstract}

\bibitem[{Davidge \& Puzia(2011)}]{davidge_stellar_2011}
Davidge, T.~J., \& Puzia, T.~H. 2011, The Astrophysical Journal, 738, 144.
\newblock
  \url{http://stacks.iop.org/0004-637X/738/i=2/a=144?key=crossref.7586b290a170d1b93a265ebf414e6e32}

\bibitem[{de~Grijs \& Bono(2014)}]{de_grijs_clustering_2014}
de~Grijs, R., \& Bono, G. 2014, The Astronomical Journal, 148, 17, aDS Bibcode:
  2014AJ....148...17D.
\newblock \url{https://ui.adsabs.harvard.edu/abs/2014AJ....148...17D}

\bibitem[{Dolphin(2002)}]{dolphin_numerical_2002}
Dolphin, A.~E. 2002, Monthly Notices of the Royal Astronomical Society, 332,
  91.
\newblock \url{http://adsabs.harvard.edu/abs/2002MNRAS.332...91D}

\bibitem[{Elmegreen(2009)}]{elmegreen_pressure_2009}
Elmegreen, B.~G. 2009, Astrophysics and Space Science, 324, 83.
\newblock \url{https://ui.adsabs.harvard.edu/abs/2009Ap&SS.324...83E/abstract}

\bibitem[{Elmegreen(2018)}]{elmegreen_two_2018}
---. 2018, The Astrophysical Journal, 869, 119, aDS Bibcode:
  2018ApJ...869..119E.
\newblock \url{https://ui.adsabs.harvard.edu/abs/2018ApJ...869..119E}

\bibitem[{Fan \& Grijs(2014)}]{fan_star_2014}
Fan, Z., \& Grijs, R.~d. 2014, The Astrophysical Journal Supplement Series,
  211, 22.
\newblock \url{https://doi.org/10.1088%2F0067-0049%2F211%2F2%2F22}

\bibitem[{Foreman-Mackey(2016)}]{corner2016}
Foreman-Mackey, D. 2016, The Journal of Open Source Software, 1, 24.
\newblock \url{https://doi.org/10.21105/joss.00024}

\bibitem[{Foreman-Mackey {et~al.}(2013)Foreman-Mackey, Hogg, Lang, \&
  Goodman}]{foreman-mackey_emcee_2013}
Foreman-Mackey, D., Hogg, D.~W., Lang, D., \& Goodman, J. 2013, Publications of
  the Astronomical Society of the Pacific, 125, 306.
\newblock \url{http://adsabs.harvard.edu/abs/2013PASP..125..306F}

\bibitem[{{Foreman-Mackey} {et~al.}(2013){Foreman-Mackey}, {Hogg}, {Lang}, \&
  {Goodman}}]{emcee2013}
{Foreman-Mackey}, D., {Hogg}, D.~W., {Lang}, D., \& {Goodman}, J. 2013, \pasp,
  125, 306

\bibitem[{Fouesneau {et~al.}(2014)Fouesneau, Johnson, Weisz, Dalcanton, Bell,
  Bianchi, Caldwell, Gouliermis, Guhathakurta, Kalirai, Larsen, Rix, Seth,
  Skillman, Williams, \& collaboration}]{fouesneau_panchromatic_2014}
Fouesneau, M., Johnson, L.~C., Weisz, D.~R., {et~al.} 2014, The Astrophysical
  Journal, 786, 117, arXiv: 1402.7264.
\newblock \url{http://arxiv.org/abs/1402.7264}

\bibitem[{Gieles(2009)}]{gieles_what_2009}
Gieles, M. 2009, Astrophysics and Space Science, 324, 299.
\newblock \url{http://adsabs.harvard.edu/abs/2009Ap%26SS.324..299G}

\bibitem[{Gieles {et~al.}(2012)Gieles, Moeckel, \& Clarke}]{gieles_all_2012}
Gieles, M., Moeckel, N., \& Clarke, C.~J. 2012, Monthly Notices of the Royal
  Astronomical Society: Letters, 426, L11, publisher: Oxford Academic.
\newblock \url{https://academic.oup.com/mnrasl/article/426/1/L11/988361}

\bibitem[{Girardi {et~al.}(2010)Girardi, Williams, Gilbert, Rosenfield,
  Dalcanton, Marigo, Boyer, Dolphin, Weisz, Melbourne, Olsen, Seth, \&
  Skillman}]{girardi_acs_2010}
Girardi, L., Williams, B.~F., Gilbert, K.~M., {et~al.} 2010, The Astrophysical
  Journal, 724, 1030.
\newblock \url{http://adsabs.harvard.edu/abs/2010ApJ...724.1030G}

\bibitem[{Goodman \& Weare(2010)}]{goodman_ensemble_2010}
Goodman, J., \& Weare, J. 2010, Communications in Applied Mathematics and
  Computational Science, 5, 65, publisher: Mathematical Sciences Publishers.
\newblock \url{https://msp.org/camcos/2010/5-1/p04.xhtml}

\bibitem[{Grudi{\'c} {et~al.}(2021)Grudi{\'c}, Kruijssen, Faucher-Gigu{\'e}re,
  Hopkins, Ma, Quataert, \& Boylan-Kolchin}]{grudic_model_2021}
Grudi{\'c}, M.~Y., Kruijssen, J. M.~D., Faucher-Gigu{\'e}re, C.-A., {et~al.}
  2021, Monthly Notices of the Royal Astronomical Society,
  doi:10.1093/mnras/stab1894.
\newblock \url{https://ui.adsabs.harvard.edu/abs/2021MNRAS.tmp.1646G}

\bibitem[{Johnson {et~al.}(2022)Johnson, Wainer, TorresVillanueva, Seth,
  Dalcanton, \& Williams}]{johnson_catalog_2022}
Johnson, L.~C., Wainer, T.~M., TorresVillanueva, E.~E., {et~al.} 2022,
  submitted to AAS Journals

\bibitem[{Johnson {et~al.}(2016)Johnson, Seth, Dalcanton, Beerman, Fouesneau,
  Lewis, Weisz, Williams, Bell, Dolphin, Larsen, Sandstrom, \&
  Skillman}]{johnson_panchromatic_2016}
Johnson, L.~C., Seth, A.~C., Dalcanton, J.~J., {et~al.} 2016, The Astrophysical
  Journal, 827, 33.
\newblock \url{https://doi.org/10.3847%2F0004-637x%2F827%2F1%2F33}

\bibitem[{Johnson {et~al.}(2017)Johnson, Seth, Dalcanton, Beerman, Fouesneau,
  Weisz, Bell, Dolphin, Sandstrom, \& Williams}]{johnson_panchromatic_2017}
---. 2017, The Astrophysical Journal, 839, 78.
\newblock \url{http://adsabs.harvard.edu/abs/2017ApJ...839...78J}

\bibitem[{Jordan {et~al.}(2007)Jordan, McLaughlin, Cote, Ferrarese, Peng, Mei,
  Villegas, Merritt, Tonry, \& West}]{jordan_acs_2007}
Jordan, A., McLaughlin, D.~E., Cote, P., {et~al.} 2007, The Astrophysical
  Journal Supplement Series, 171, 101, arXiv: astro-ph/0702496.
\newblock \url{http://arxiv.org/abs/astro-ph/0702496}

\bibitem[{Kass \& Raftery(1995)}]{kass_bayes_1995}
Kass, R.~E., \& Raftery, A.~E. 1995, Journal of the American Statistical
  Association, 90, 773, publisher: [American Statistical Association, Taylor \&
  Francis, Ltd.].
\newblock \url{https://www.jstor.org/stable/2291091}

\bibitem[{Kelly(2007)}]{kelly_aspects_2007}
Kelly, B.~C. 2007, The Astrophysical Journal, 665, 1489, aDS Bibcode:
  2007ApJ...665.1489K.
\newblock \url{https://ui.adsabs.harvard.edu/abs/2007ApJ...665.1489K}

\bibitem[{King(1962)}]{king_structure_1962}
King, I. 1962, The Astronomical Journal, 67, 471.
\newblock \url{http://adsabs.harvard.edu/abs/1962AJ.....67..471K}

\bibitem[{Kroupa(2001)}]{kroupa_variation_2001}
Kroupa, P. 2001, Monthly Notices of the Royal Astronomical Society, 322, 231,
  aDS Bibcode: 2001MNRAS.322..231K.
\newblock \url{https://ui.adsabs.harvard.edu/abs/2001MNRAS.322..231K}

\bibitem[{Kruijssen(2012)}]{kruijssen_fraction_2012}
Kruijssen, J. M.~D. 2012, Monthly Notices of the Royal Astronomical Society,
  426, 3008, aDS Bibcode: 2012MNRAS.426.3008K.
\newblock \url{https://ui.adsabs.harvard.edu/abs/2012MNRAS.426.3008K}

\bibitem[{Kruijssen {et~al.}(2019)Kruijssen, Pfeffer, Reina-Campos, Crain, \&
  Bastian}]{kruijssen_formation_2019}
Kruijssen, J. M.~D., Pfeffer, J.~L., Reina-Campos, M., Crain, R.~A., \&
  Bastian, N. 2019, Monthly Notices of the Royal Astronomical Society, 486,
  3180, aDS Bibcode: 2019MNRAS.486.3180K.
\newblock \url{https://ui.adsabs.harvard.edu/abs/2019MNRAS.486.3180K}

\bibitem[{Krumholz {et~al.}(2015)Krumholz, Fumagalli, da~Silva, Rendahl, \&
  Parra}]{krumholz_slug_2015}
Krumholz, M.~R., Fumagalli, M., da~Silva, R.~L., Rendahl, T., \& Parra, J.
  2015, Monthly Notices of the Royal Astronomical Society, 452, 1447.
\newblock
  \url{https://academic.oup.com/mnras/article-lookup/doi/10.1093/mnras/stv1374}

\bibitem[{Krumholz {et~al.}(2019)Krumholz, McKee, \&
  Bland-Hawthorn}]{krumholz_star_2019}
Krumholz, M.~R., McKee, C.~F., \& Bland-Hawthorn, J. 2019, Annual Review of
  Astronomy and Astrophysics, 57, 227.
\newblock \url{http://adsabs.harvard.edu/abs/2019ARA%26A..57..227K}

\bibitem[{Larsen(2009)}]{larsen_mass_2009}
Larsen, S.~S. 2009, Astronomy and Astrophysics, 494, 539.
\newblock \url{http://adsabs.harvard.edu/abs/2009A%26A...494..539L}

\bibitem[{Leroy {et~al.}(2008)Leroy, Walter, Brinks, Bigiel, de~Blok, Madore,
  \& Thornley}]{leroy_star_2008}
Leroy, A.~K., Walter, F., Brinks, E., {et~al.} 2008, The Astronomical Journal,
  136, 2782, aDS Bibcode: 2008AJ....136.2782L.
\newblock \url{https://ui.adsabs.harvard.edu/abs/2008AJ....136.2782L}

\bibitem[{Lieberz \& Kroupa(2017)}]{lieberz_origin_2017}
Lieberz, P., \& Kroupa, P. 2017, Monthly Notices of the Royal Astronomical
  Society, 465, 3775, aDS Bibcode: 2017MNRAS.465.3775L.
\newblock \url{https://ui.adsabs.harvard.edu/abs/2017MNRAS.465.3775L}

\bibitem[{Longmore {et~al.}(2014)Longmore, Kruijssen, Bastian, Bally,
  Rathborne, Testi, Stolte, Dale, Bressert, \& Alves}]{longmore_formation_2014}
Longmore, S.~N., Kruijssen, J. M.~D., Bastian, N., {et~al.} 2014, conference
  Name: Protostars and Planets VI Pages: 291 ADS Bibcode: 2014prpl.conf..291L.
\newblock \url{https://ui.adsabs.harvard.edu/abs/2014prpl.conf..291L}

\bibitem[{Marigo {et~al.}(2008)Marigo, Girardi, Bressan, Groenewegen, Silva, \&
  Granato}]{marigo_evolution_2008}
Marigo, P., Girardi, L., Bressan, A., {et~al.} 2008, Astronomy and
  Astrophysics, Volume 482, Issue 3, 2008, pp.883-905, 482, 883.
\newblock
  \url{https://ui.adsabs.harvard.edu/abs/2008A%26A...482..883M/abstract}

\bibitem[{Messa {et~al.}(2018)Messa, Adamo, {\"O}stlin, Calzetti, Grasha,
  Grebel, Shabani, Chandar, Dale, Dobbs, Elmegreen, Fumagalli, Gouliermis, Kim,
  Smith, Thilker, Tosi, Ubeda, Walterbos, Whitmore, Fedorenko, Mahadevan,
  Andrews, Bright, Cook, Kahre, Nair, Pellerin, Ryon, Ahmad, Beale, Brown,
  Clarkson, Guidarelli, Parziale, Turner, \& Weber}]{messa_young_2018}
Messa, M., Adamo, A., {\"O}stlin, G., {et~al.} 2018, Monthly Notices of the
  Royal Astronomical Society, 473, 996.
\newblock \url{http://adsabs.harvard.edu/abs/2018MNRAS.473..996M}

\bibitem[{Messa {et~al.}(2021)Messa, Calzetti, Adamo, Grasha, Johnson, Sabbi,
  Smith, Bajaj, Finn, \& Lin}]{messa_looking_2021}
Messa, M., Calzetti, D., Adamo, A., {et~al.} 2021, The Astrophysical Journal,
  909, 121, aDS Bibcode: 2021ApJ...909..121M.
\newblock \url{https://ui.adsabs.harvard.edu/abs/2021ApJ...909..121M}

\bibitem[{Mok {et~al.}(2019)Mok, Chandar, \& Fall}]{mok_constraints_2019}
Mok, A., Chandar, R., \& Fall, S.~M. 2019, The Astrophysical Journal, 872, 93,
  aDS Bibcode: 2019ApJ...872...93M.
\newblock \url{https://ui.adsabs.harvard.edu/abs/2019ApJ...872...93M}

\bibitem[{Portegies~Zwart {et~al.}(2010)Portegies~Zwart, McMillan, \&
  Gieles}]{portegies_zwart_young_2010}
Portegies~Zwart, S.~F., McMillan, S. L.~W., \& Gieles, M. 2010, Annual Review
  of Astronomy and Astrophysics, vol. 48, p.431-493, 48, 431.
\newblock
  \url{https://ui.adsabs.harvard.edu/abs/2010ARA%26A..48..431P/abstract}

\bibitem[{Reina-Campos \& Kruijssen(2017)}]{reina-campos_unified_2017}
Reina-Campos, M., \& Kruijssen, J. M.~D. 2017, Monthly Notices of the Royal
  Astronomical Society, 469, 1282.
\newblock \url{http://adsabs.harvard.edu/abs/2017MNRAS.469.1282R}

\bibitem[{San~Roman {et~al.}(2010)San~Roman, Sarajedini, \&
  Aparicio}]{san_roman_photometric_2010}
San~Roman, I., Sarajedini, A., \& Aparicio, A. 2010, The Astrophysical Journal,
  720, 1674.
\newblock \url{https://doi.org/10.1088%2F0004-637x%2F720%2F2%2F1674}

\bibitem[{Schechter(1976)}]{schechter_analytic_1976}
Schechter, P. 1976, The Astrophysical Journal, 203, 297.
\newblock \url{http://adsabs.harvard.edu/abs/1976ApJ...203..297S}

\bibitem[{U {et~al.}(2009)U, Urbaneja, Kudritzki, Jacobs, Bresolin, \&
  Przybilla}]{u_new_2009}
U, V., Urbaneja, M.~A., Kudritzki, R.-P., {et~al.} 2009, The Astrophysical
  Journal, 704, 1120, aDS Bibcode: 2009ApJ...704.1120U.
\newblock \url{https://ui.adsabs.harvard.edu/abs/2009ApJ...704.1120U}

\bibitem[{Utomo {et~al.}(2019)Utomo, Blitz, \& Falgarone}]{utomo_origin_2019}
Utomo, D., Blitz, L., \& Falgarone, E. 2019, The Astrophysical Journal, 871,
  17, aDS Bibcode: 2019ApJ...871...17U.
\newblock \url{https://ui.adsabs.harvard.edu/abs/2019ApJ...871...17U}

\bibitem[{Weisz {et~al.}(2013)Weisz, Fouesneau, Hogg, Rix, Dolphin, Dalcanton,
  Foreman-Mackey, Lang, Johnson, Beerman, Bell, Gordon, Gouliermis, Kalirai,
  Skillman, \& Williams}]{weisz_panchromatic_2013}
Weisz, D.~R., Fouesneau, M., Hogg, D.~W., {et~al.} 2013, The Astrophysical
  Journal, 762, 123.
\newblock \url{http://adsabs.harvard.edu/abs/2013ApJ...762..123W}

\bibitem[{Weisz {et~al.}(2015)Weisz, Johnson, Foreman-Mackey, Dolphin, Beerman,
  Williams, Dalcanton, Rix, Hogg, Fouesneau, Johnson, Bell, Boyer, Gouliermis,
  Guhathakurta, Kalirai, Lewis, Seth, \& Skillman}]{weisz_high-mass_2015}
Weisz, D.~R., Johnson, L.~C., Foreman-Mackey, D., {et~al.} 2015, The
  Astrophysical Journal, 806, 198.
\newblock \url{http://adsabs.harvard.edu/abs/2015ApJ...806..198W}

\bibitem[{Whitmore {et~al.}(2020)Whitmore, Chandar, Lee, Ubeda, Adamo, Aloisi,
  Calzetti, Cignoni, Cook, Dale, Elmegreen, Gouliermis, Grebel, Grasha,
  Johnson, Kim, Sacchi, Smith, Tosi, \& Wofford}]{whitmore_legus_2020}
Whitmore, B.~C., Chandar, R., Lee, J., {et~al.} 2020, The Astrophysical
  Journal, 889, 154, aDS Bibcode: 2020ApJ...889..154W.
\newblock \url{https://ui.adsabs.harvard.edu/abs/2020ApJ...889..154W}

\bibitem[{Williams {et~al.}(2009)Williams, Dalcanton, Dolphin, Holtzman, \&
  Sarajedini}]{williams_detection_2009}
Williams, B.~F., Dalcanton, J.~J., Dolphin, A.~E., Holtzman, J., \& Sarajedini,
  A. 2009, The Astrophysical Journal Letters, 695, L15.
\newblock \url{http://adsabs.harvard.edu/abs/2009ApJ...695L..15W}

\bibitem[{Williams {et~al.}(2021)Williams, Durbin, Dalcanton, Lang, Girardi,
  Smercina, Dolphin, Weisz, Choi, Bell, Rosolowsky, Skillman, Koch, Lindberg,
  Hagen, Gordon, Seth, Gilbert, Guhathakurta, Lauer, \&
  Bianchi}]{williams_panchromatic_2021}
Williams, B.~F., Durbin, M.~J., Dalcanton, J.~J., {et~al.} 2021, arXiv
  e-prints, 2101, arXiv:2101.01293.
\newblock \url{http://adsabs.harvard.edu/abs/2021arXiv210101293W}

\bibitem[{Zhang \& Fall(1999)}]{zhang_mass_1999}
Zhang, Q., \& Fall, S.~M. 1999, The Astrophysical Journal Letters, 527, L81.
\newblock \url{http://adsabs.harvard.edu/abs/1999ApJ...527L..81Z}

\end{thebibliography}

\end{document}